\documentclass{article}
\usepackage{amsthm,amsmath,amssymb,bm}
\usepackage{mathrsfs}
\usepackage{float} 
\usepackage{graphicx}
\usepackage{footnote}
\usepackage{amsthm,amsmath,amssymb,bm}
\usepackage{mathrsfs}
\usepackage{float} 
\usepackage{graphicx}
\usepackage{footnote}
\usepackage{fdsymbol}
\usepackage{multirow}
\usepackage{cite}
\usepackage[authoryear]{natbib}
\usepackage{algorithm}
\usepackage{algorithmic}
\usepackage{threeparttable}
\usepackage{lscape}

\usepackage{amsthm,amsmath,amssymb,bm}
\usepackage{mathrsfs}
\usepackage{float} 
\usepackage{graphicx}
\usepackage{subfigure}
\usepackage{float}
\usepackage{footnote}
\usepackage{cite}
\usepackage{soul}
\usepackage{authblk}
\usepackage{color,xcolor}
\usepackage{amsmath,amsfonts,amssymb,amsthm,booktabs,color,epsfig,graphicx,hyperref,url}
\usepackage{mathrsfs}
\usepackage{float} 
\usepackage{graphicx}
\usepackage{footnote}
\usepackage{fdsymbol}
\usepackage{cite}
\usepackage{multirow}
\usepackage{subfigure}
\usepackage{algorithm}
\usepackage{algorithmic}
\usepackage{soul}
\usepackage{color,xcolor}
\soulregister{\citet}7
\soulregister{\romannumeral}7
\sethlcolor{yellow}

\begin{document}
	
	\title{Weighted Euclidean balancing for a matrix exposure in estimating causal effect}
	\author[1]{Juan Chen}
	\author[1]{Yingchun Zhou \thanks{Corresponding author:  yczhou@stat.ecnu.edu.cn}}
	\affil[1]{Key Laboratory of Advanced Theory and Application in Statistics
		and Data Science-MOE, School of Statistics, East China Normal University.}
	\date{}
	
	\maketitle
	\begin{abstract}
		In many scientific fields such as biology, psychology and sociology, there is an increasing interest in estimating the causal effect of a matrix exposure on an outcome.  Covariate balancing is crucial in causal inference and both exact balancing and approximate balancing methods have been proposed in the past decades. However, due to the large number of constraints, it is difficult to achieve exact balance or to select the threshold parameters for approximate balancing methods when the treatment is a matrix. To meet these challenges, we propose the weighted Euclidean balancing method, which approximately balance covariates from an overall perspective. This method is also applicable to high-dimensional covariates scenario. Both parametric and nonparametric methods are proposed to estimate the causal effect of matrix treatment and theoretical properties of the two estimations are provided. Furthermore, the simulation results show that the proposed method outperforms other methods in various cases. Finally, the method is applied to investigating the causal relationship between children's participation in various training courses and their IQ. The results show that the duration of attending hands-on practice courses for children at 6-9 years old has a siginificantly positive impact on children's IQ.
	\end{abstract}
	\textbf{Keywords}: causal inference, matrix treatment, weighting methods, overall imbalance, observational study. 
	
	\newpage
	
	\section{Introduction}
	For decades, causal inference has been widely used in many fields, such as biology, psychology and economics, etc. Most of the current research is based on univariate treatment (binary, multivalued or continuous treatment)  (\citet{imai2014covariate}; \citet{zhu2015boosting}; \citet{fong2018covariate}; \citet{zubizarreta2015stable}; \citet{chan2016globally}; \citet{xiong2017treatment}; \citet{yiu2018covariate}; \citet{dong2021regression}; \citet{hsu2022counterfactual}). However, one may be interested in the causal effect of a matrix treatment. For example, in studying the impact of children's participation in training courses on children's intelligence (measured by IQ), the exposure is a matrix, whose rows represent different age groups, columns represent the types of trainging courses and each element represents the number of hours of class per week. The existing methods are not suitable for matrix exposure and there have been few research on this. Therefore, the goal of this paper is to develop a new method to estimate the causal effect function for matrix exposure.
	\noindent
	
	To estimate causal effects in observational studies, it is common to use propensity scores (\citet{1983Central}; \citet{imbens2000role}; \citet{imai2004causal}). There are several classes of propensity score-based methods, such as matching, weighting and subclassification, that have become part of applied researchers' standard toolkit across many scientific displines (\citet{2004Stratification}; \citet{rubin2006matched}). In this article we focus on the weighting method. 
	\noindent
	
	In the past decade, various weighting methods have been proposed to balance covariates in the estimation procedure (\citet{hainmueller2012entropy}; \citet{imai2004causal}; \citet{vegetabile2020optimally}). The key idea of these methods is to estimate propensity score ( \citet{rosenbaum1984reducing}; \citet{rosenbaum1985constructing}; \citet{robins2000marginal}; \citet{hirano2004propensity}).
	%Rosenbaum and Rubin, 1983, 1984, 1985; Robins et al., 2000; Hirano et al., 2003). 
	When using the parametric method to model the propensity score, the estimation bias of the causal effect will be large if the model is mis-specified. Therefore, some nonparametric methods for estimating the propensity score have been proposed, such as the kernel density estimation (\citet{robbins2020robust}). In addition, in recent years, some studies have used optimized weighting methods to directly optimize the balance of covariates (\citet{hainmueller2012entropy}; \citet{imai2004causal}; \citet{vegetabile2020optimally}).
	%(Hainmueller 2012; Imai and Ratkovic 2014; Vegetabile et al., 2020). 
	These methods avoid the direct construction of the propensity scores, therefore the obtained estimate achieves higher robustness. One of the methods, the entropy balancing method, has been established as being doubly robust, in that a consistent estimate can still be obtained when one of the two models, either the treatment assignment model or the outcome model, is correctly specified (\citet{zhao2017entropy}).
	%(Zhao and Percival 2017). 
	Furthermore, this method can be easily implemented by solving a convex optimization problem. Here we extend the entropy balancing method to the matrix treatment scenario to balance the covariates.
	\noindent
	
	The methods mentioned above assume that all balancing conditions hold exactly, that is, they are exact balancing methods. However, the balancing conditions cannot hold exactly when the dimension of covariate or treatment is high as there will be too many equations to hold simultaneously. For matrix treatment, it is even more difficult for the balancing conditions to hold exactly. To meet this challenge, literatures have shown that approximate balance can trade bias for variance whereas exact balance cannot and the former works well in practice in both low- and high-dimensional settings (\citet{wang2020minimal}; \citet{2018Approximate}). The potential limitations of the existing approximate balancing methods are that they directly control univariate imbalance, which cannot guarantee the overall balance especially in the high-dimensional scenario. Besides, there is no principled way to select the threshold parameters simultaneously in practice. Another potential issue of the univariate approximate balancing methods is that it is difficult to handle high-dimensional constraints since the theoretical results require that the number of the balancing constraints should be much smaller than the sample size(\citet{wang2020minimal}).
	\noindent

	To alleviate the limitations of univariate balancing methods, we propose an overall balancing approach, which is called Weighted Euclidean balancing method. The weight is obtained by optimizing the entropy function subject to a single inequality constraint, hence the issue of tuning multiple threshold parameters in univariate balancing methods is solved. The Weighted Euclidean distance is defined to measure the overall imbalance and a sufficient small value of the distance suggests that the covariates are approximately balanced from the overall perspective. Moreover, we propose an algorithm to deal with high-dimensional constraints, so that the proposed method is not restrictive to the low-dimensional setting.
	\noindent
	
	The main contributions of the paper are summarized as follows. First, an overall balancing method (Weighted Euclidean balancing method) is proposed, which extends the binary univariate entropy approximate balancing method to the matrix treatment scenario. Unlike univariate approximate balancing method, the Weighted Euclidean balancing method controls the imbalance  from the overall perspective. Moreover, to the best of our knowledge, it is the first time that matrix treatment is studied by weighting method in causal inference literature. Second, both parametric and nonparametric causal effect estimation methods for matrix treatment are proposed. Under the parametric framework, a weighted optimization estimation is defined and its theoretical properties are provided. Under the nonparametric framework, B-splines are used to approximate the causal effect function and the convergence rate of the estimation is provided. Third, the proposed method is applied to explore the impact of children's participation in training courses on their IQ and meaningful results are obtained.
	\noindent
	
	The remainder of this article is organized as follows: In Section 2, the preliminaries are introduced. In Section 3, the Weighted Euclidean balancing method (WEBM) is proposed. In Section 4, the theoretical properties of the WEBM method are shown. In section 5, a numerical simulation is performed to evaluate the performance of the WEBM method under finite samples. In Section 6, the WEBM method is applied to analyze a real problem. The conclusions and discussions are summarized in Section 7.

	\section{Preliminary}
	\subsection{Notation and assumptions}
	Suppose an independent and identically distributed sample $(\mathbf{Z}_1,\dots,\mathbf{Z}_n)$ is observed, where the support of $\mathbf{Z} = (\mathbf{T},\mathbf{X},Y)$ is $\mathcal{Z}=(\mathcal{T} \times \mathcal{X} \times \mathcal{Y})$. Here $\mathbf{T} \in R^{p\times q}$ denotes a matrix exposure, $\mathbf{X} \in R^J$ denotes a vector of covariates, and $Y \in R$ denotes the observed outcome. Since the causal effect is characterized by potential outcome notion, let $Y(\textbf{t})$ for all $\textbf{t} \in \mathcal{T}$ denotes the potential outcome that would be observed under treatment level $\mathbf{t}$, i.e. $Y = Y(\textbf{t})$ if $\textbf{T}= \textbf{t}$.  
	
	\noindent 
	
	In this paper, our goal is to estimate the causal effect function $\mathbb{E}(Y(\mathbf{t}))$, which is defined in terms of potential outcomes that are not directly observed. Therefore, three assumptions that commonly employed for indentification are made (\citet{hirano2004propensity}; \citet{imai2004causal}).\\
	%(Hirano and Imbens, 2004; Imai and van Dyk, 2004):
	
	\noindent
	\textbf{Assumption 1 (Ignorability)}:\\
	\\
	$\textbf{T}_i \perp Y_i(\textbf{t}) \mid \textbf{X}_i$, which implies that the set of observed pre-treatment covariates $\mathbf{X}_i$, is sufficiently rich such that it includes all confounders , i.e. there is no unmeasured confounding.
	\\
	
	\noindent
	\textbf{Assumption 2 (Positivity)}:\\
	\\
	$f_{\textbf{T} \mid \textbf{X}}(\textbf{T}_i = \textbf{t} \mid \textbf{X}_i ) > 0$ for all $\textbf{t} \in \mathcal{T}$, which means that treatment is not assigned deterministically.\\
	%(Imbens, 2000).
	
	\noindent
	\textbf{Assumption 3 (SUTVA)}:\\
	\\
	Assume that there is no interference among the units, which means that each individual's outcome depends only on their own level of treatment intensity.
	\\
	
	Under the above assumptions, we first define the stabilized weight as\\
	\begin{equation*}
		w_i = \frac{f(\textbf{T}_i)}{f(\textbf{T}_i \mid \textbf{X}_i)},
	\end{equation*}
	then one can estimate the causal effect function based on the stabilized weight with observational data.
	
	\subsection{Exact entropy balancing and approximate entropy balancing for matrix exposure}
	The entropy balancing method (\citet{hainmueller2012entropy})
	%(Hainmueller 2012) 
	is used to determine the optimal weight for inferring causal effects. It has been used for univariate treatment and here this method is extended to matrix exposure and to balance covariates approximately.	
	\noindent
	
	Note that the stabilized weight \\
	\begin{equation}
		w_i = \frac{f(\textbf{T}_i)}{f(\textbf{T}_i \mid \textbf{X}_i)}
	\end{equation}
	satisfies the following conditions for any suitable functions $u(\mathbf{T})$ and $v(\mathbf{X})$:
	\begin{equation}
		\begin{split}
			\mathbb{E}(w_iu(\mathbf{T}_i)v(\mathbf{X}_i)) = & \iint \frac{f(\textbf{T}_i)}{f(\textbf{T}_i \mid \textbf{X}_i)}u(\mathbf{T}_i)v(\mathbf{X}_i)f(\mathbf{T}_i, \textbf{X}_i)d\textbf{T}_id\textbf{X}_i \\
			&=\int \lbrace \frac{f(\textbf{T}_i)}{f(\textbf{T}_i \mid \textbf{X}_i)}u(\textbf{T}_i)f(\textbf{T}_i \mid \textbf{X}_i)d\textbf{T}_i \rbrace v(\textbf{X}_i)f(\textbf{X}_i)d\textbf{X}_i \\
			&=\mathbb{E}(u(\mathbf{T}_i))\mathbb{E}(v(\mathbf{X}_i))
		\end{split}
	\end{equation}
	Besides, it also satisfies that 
	\begin{equation}
		\mathbb{E}(w_i) = \iint \frac{f(\textbf{T}_i)}{f(\textbf{T}_i \mid \textbf{X}_i)}f(\textbf{T}_i, \textbf{X}_i)d\textbf{T}_id\textbf{X}_i = 1.
	\end{equation}
	However, Equation (2) implies an infinite number of moment conditions, which is impossible to solve with a finite sample of observations. Hence, the finite dimensional sieve space is considered to approximate the infinite dimensional function space. Specifically, let 
	\begin{gather*}
		u_{K1}(\mathbf{T}) = (u_{K1,1}(\mathbf{T}), u_{K1,2}(\mathbf{T}),\dots, u_{K1,K1}(\mathbf{T}))^{'}, \\
		v_{K2}(\mathbf{X}) = (v_{K2,1}(\mathbf{X}), v_{K2,2}(\mathbf{X}), \dots, v_{K2,K2}(\mathbf{X}))^{'} 
	\end{gather*}
	denote the known basis functions, then 
	\begin{equation}
		\mathbb{E}(w_i u_{K1}(\mathbf{T}_i) v_{K2}(\mathbf{X}_i)^{'}) = \mathbb{E}(u_{K1}(\mathbf{T}_i))\mathbb{E}(v_{K2}(\mathbf{X}_i)^{'}). 
	\end{equation}
	
	\noindent
	
	In practice, the covariate balancing conditions given in Equation (4) cannot hold exactly with high dimensional covariates or treatments. It is even more difficult to hold exactly for matrix exposure. To overcome this difficulty, approximate balance is considered rather than exact balance, which has been demonstrated to work well in practice in both low- and high-dimensional settings (\citet{2015Stable}; \citet{2018Approximate}; \citet{wang2020minimal}). Specifically, the balancing weights that approximately satisfy the conditions in Equation (4) are the global minimum of the following optimization problem:
	\begin{gather}
		\text{min}_\mathbf{w} \sum_{i=1}^{n}w_ilog(w_i) \notag                                   
	\end{gather}
	s.t.
	\begin{gather}
		\mid	\frac{1}{n}\sum_{i=1}^{n}w_i u_{K1,l}(\mathbf{T}_i)v_{K2,\tilde{l}}(\mathbf{X}_i) - (\frac{1}{n}\sum_{i=1}^{n} u_{K1,l}(\mathbf{T}_i)) (\frac{1}{n} \sum_{i=1}^{n}v_{K2,\tilde{l}}(\mathbf{X}_i)) \mid \leq \delta_{l,\tilde{l}},
	\end{gather}
	where $u_{K1,l}(\mathbf{T}_i)$ and $v_{K2,\tilde{l}}(\mathbf{X}_i)$ denote the $l$th and $\tilde{l}$th components of $u_{K1}(\mathbf{T}_i)$ and $v_{K2}(\mathbf{X}_i)$, respectively.
	Let $m_K(\mathbf{T}_i, \mathbf{X}_i) = \text{vec}(\frac{1}{n} u_{K1}(\mathbf{T}_i)v_{K2}(\mathbf{X}_i)^{'})$ and $\bar{m}_K = \text{vec} (\frac{1}{n}\bar{u}_{K1}\bar{v}_{K2}^{'})$  denote two column vectors with dimension $K$, where $K= K1 K2 $, the $l$th and $\tilde{l}$th components of $\bar{u}_{K1}$ and $\bar{v}_{K2}$ are defined as
	\begin{equation}
		\bar{u}_{K1,l} = \frac{1}{n}\sum_{i=1}^{n} u_{K1,l}(\mathbf{T}_i) \ \text{and} \ \bar{v}_{K2,\tilde{l}} = \frac{1}{n}\sum_{i=1}^{n} v_{K2,\tilde{l}}(\mathbf{X}_i), 
	\end{equation}
	then condition (5) is equivalent to 
	\begin{gather}
		\text{min}_w \sum_{i=1}^{n}w_ilog(w_i)  \notag                                   
	\end{gather}
	s.t.
	\begin{gather}
		\mid	\sum_{i=1}^{n}w_i m_{K,k}(\mathbf{T}_i, \mathbf{X}_i)  -  n\bar{m}_{K,k} \mid \leq \delta_{k}, \ k= 1,\dots,K.
	\end{gather}
	\noindent
	However, there is a large number of tuning parameters $(\delta_1,\dots,\delta_K)$ which is very time-consuming to determine and there is lack of guideline on tuning these parameters simultaneously in practice.

	\section{Methodology}
	Due to the potential issues of univariate approximate balancing methods, the weighted Euclidean balancing method is proposed in this section, whose key idea is to control the overall imbalance in the optimization problem (7). 
	\noindent

	\subsection{Weighted Euclidean balancing method}
	%To meet the challenge of the univariate approximate balancing method, we propose the weighted Euclidean balancing (WEB) method, whose key idea is to directly control overall imbalance in the optimization problem (7).
	Define the following weighted Euclidean imbalance measure (WEIM) as a weighted version of the squared Euclidean distance:
	\begin{equation}
		\text{WEIM} = \sum_{k=1}^{K} \lbrace \lambda_k^2 [\sum_{i=1}^{n} w_i (m_{K,k}(\mathbf{T}_i,\mathbf{X}_i)-\bar{m}_{K,k})]^2 \rbrace.
	\end{equation}
	The weighted Euclidean balancing obtains the balancing weights that approximately satisfy the condition (4) by solving the following convex optimization problem:
	\begin{gather}
		\text{min}_w \sum_{i=1}^{n}w_ilog(w_i)  \notag                                   
	\end{gather}
	s.t.
	\begin{gather}
		\sum_{k=1}^{K} \lbrace \lambda_k^2 [\sum_{i=1}^{n} w_i (m_{K,k}(\mathbf{T}_i,\mathbf{X}_i)-\bar{m}_{K,k})]^2 \rbrace \leq \delta,
	\end{gather}
	where $(\lambda_1,\dots, \lambda_K)$ is a pre-sepecified weight vector and $\delta \geq 0$ is a threshold parameter. Assume that condition (3) holds exactly, whose sample condition is $\frac{1}{n} \sum_{i=1}^{n} w_i = 1$.
	\noindent
	
	%The pre-specified vector  $(\lambda_1,\dots, \lambda_K)$ reflects the importance of the basis functions
	%and their interactions. If we set $\lambda_j = 1, j= 1,\dots,K$, then we only consider the Euclidean balancing method, which means the effects of high-order interactions of the basis functions may be as large as the main effects. However, models in which effects tend to be smaller for higher order interactions, like the one used in \citet{2021Multilevel}, tend to be more plausible. To encode the classical assumption that the coefficients corresponding to the main effects are larger than the effects of first-order interactions, which are larger than those of second-order interactions, and so on, we define $\lambda_j , j= 1,\dots,K$ as a decreasing function of the interaction order. Taking $\lambda_j = 0$ corresponds to assuming away any effect for $m_{K,j}(\mathbf{T},\mathbf{X})$, and taking $\lambda_j = \infty$ corresponds to assuming no limits on that effect and therefore requires exact balance in $m_{K,j}(\mathbf{T},\mathbf{X})$.
	\noindent

	Note that univariate exact balance is equivalent to the overall exact balance, in the sense that $\sum_{i=1}^{n} w_i (m_{K,k}(\mathbf{T}_i,\mathbf{X}_i)-\bar{m}_{K,k})=0, k= 1,\dots,K$  is equivalent to $\text{WEIM}=0$. However, the univariate approximate balance does not imply the overall approximate balance since it is possible that $\sum_{i=1}^{n} w_i (m_{K,k}(\mathbf{T}_i,\mathbf{X}_i)-\bar{m}_{K,k}), k= 1,\dots,K$ is small while the $\text{WEIM}$ is large.
	\noindent
	
	The pre-specified vector  $(\lambda_1,\dots, \lambda_K)$ reflects the importance of each univariate constraint. In this paper, we set $\lambda_k= \sigma_k^{-1}$, where $\sigma_k^2$ is the variance of $m_{K,k}(\mathbf{T},\mathbf{X})$. Since problem (9) is difficult to solve numerically, its dual problem is considered here, which can be solved by numerically efficient algorithms. Theorem 1 provides the dual formulation of problem (9) as an unconstrained problem.
	\\
	
	\noindent
	\textbf{Theorem 1}. \ Assume that $\text{max}_i (\text{max}_k \mid \lambda_km_{K,k}(\mathbf{T}_i,\mathbf{X}_i) \mid) < \infty$, the dual of problem (9) is equivalent to the following unconstrained problem 
	\begin{equation}
		\text{min}_{\mathbf{\theta} \in R^K}  \sum_{i=1}^{n} \text{exp}( \sum_{k=1}^{K} \theta_j \lambda_j (m_{K,k}(\mathbf{T}_i,\mathbf{X}_i)-\bar{m}_{K,k}))  + \sqrt{\delta}\mid\mid \theta \mid\mid_2,
	\end{equation}
	%where $f^{*}(x) = sup_t(xt-xlog(x))$ is the conjugate function of $f(x) = xlog(x)$.
	and the primal solution $\hat{w}_i$ is given by 
	\begin{equation}
		\hat{w}_i = \text{exp} \lbrace \sum_{k=1}^{K} \hat{\theta}_k \lambda_k (m_{K,k}(\mathbf{T}_i,\mathbf{X}_i)-\bar{m}_{K,k}) -1\rbrace , \ i=1,\dots,n,
	\end{equation}
	where $\hat{\theta}$ is the solution to the dual optimization problem (10). 
	\noindent
	The proof of Theorem 1 is in Appendix A.1. 
	
	\subsubsection*{Selection of tuning parameter}
	Another practical issue that arises with weighted Euclidean weights is how to choose the degree of approximate balance $\delta$. A tuning algorithm is proposed as follows. First, determine a range of positive values $\mathcal{D}$ for $\delta$, then the optimal value of $\delta$ is selected by the following algorithm.
	\\
	
	\noindent
	\textbf{ Algorithm 1.} Selection of $\delta$.\\
	\noindent
	
	For each $\delta \in \mathcal{D}$,
	\begin{enumerate}
		\item[1. ] Compute the dual parameters $\hat{\theta}$ by solving the dual problem (10);
		\item[2. ] Compute the estimated weights $\hat{w}_i$ using equation (11);
		\item[3. ] Calculate $\text{WEIM}$ in (8) using $\hat{w}_i$;
	\end{enumerate}
	\noindent
	
	Output $\delta^*$ that minimizes $\text{WEIM}$.
	
	\subsubsection*{Weighted Euclidean balancing with high-dimensional covariates}
	In the high-dimensional or ultra high-dimensional covariate setting with $K$ relatively large compared to $n$ or $K>>n$, it becomes difficult to control the overall imbalance using the Weighted Euclidean balancing method. To meet this challenge, we propose an algorithm to select a small subset of the covariates in the sparse setting. Specifically, consider $v_{K2}(\mathbf{X}) = (1, \mathbf{X})$ in the high-dimensional setting. Let $Bcor_j$ be the ball correlation (\citet{pan2020ball}) between $X_j$ and $\mathbf{T}, j=1,\dots,L$. Rank $X_1, \dots,X_L$ as $X_{(1)},\dots,X_{(L)}$ such that $X_{(1)}$ has the largest $Bcor$ value, $X_{(2)}$ has the second largest $Bcor$ value, and so forth. The covariates are added successively according to the rank of ball correlation until there is a break point of WEIM's, and the set added before the break point appears is the target set. The key idea hings on WEIM, which represents the contribution of the $j$ most imbalanced covariates to the overall imbalance after WEBM weighting.  If WEIM remains stable as j inceases, it indicates that the overall imbalance can be controlled. However, if there is a break point at Step j, it implies that adding the  jth covariates greatly inceases WEIM, which is harmful to the control of the overall imbalance. Therefore, the algorithm should be stopped and print the outputs at Step $j-1$. Specifically, procedures to select the subset of covariates are given by the following algorithm.\\

	\noindent
	\textbf{ Algorithm 2.} Subset selection of covariates in the high dimensional case.\\
	\noindent
	
	For each $j \in \lbrace 1,\dots, L\rbrace$,
	\begin{enumerate}
		\item[ ] compute the estimated weights $\hat{w}_i^{(j)}$ using $v_{K2}(\mathbf{X}) = (1,X_{(1)},\dots, X_{(j)})$;
		\item[  ] calculate $\text{WEIM}^{(j)}$ in (8) using $\hat{w}_i^{(j)}$;
		\item[ ] add $(j,\text{WEIM}^{(j)})$ to the $x-y$ plot and observe whether there is a break point at $(j,\text{WEIM}^{(j)})$:
		\begin{itemize}
			\item [ ] If no, let $j=j+1$;
			\item [ ] If yes, stop and output $L_0 = j-1$.
		\end{itemize}
	\end{enumerate}
	\noindent
	
	The selected subset of the covariates is $(X_{(1)},\dots, X_{(L_0)})$.

	\subsection{Causal effect estimation}
	
	In this subsection, both parametric and nonparametric approaches are developed to estimate the causal effect function. A weighted optimization estimation is defined under the parametric framework and broadcasted nonparametric tensor regression method (\citet{zhou2020broadcasted}) is used to estimate the causal effect function under the nonparametric framework.
	\noindent
	\subsubsection{Parametric approach}
	The causal effect function is parametrized as $ s(\mathbf{t};\mathbf{\beta})$, assume that it has a unique solution  $\beta^*$ defined as
	\begin{equation}
		\beta^* = \text{agrmin}_\beta \int_{\mathcal{T}} \mathbb{E}[Y(\textbf{t})- s(\mathbf{t};\mathbf{\beta}) ]^2f_\textbf{T}(\textbf{t})d\textbf{t}.
	\end{equation}
	
	The difficulty of solving Equation (12) is that the potential outcome $Y(\textbf{t})$ is not observed for all $\textbf{t}$. Hence, Proposition 1 is proposed to connect the potential outcome with the observed outcome. \\
	
	\noindent
	\textbf{Proposition 1} \ Under Assumption 1, it can be shown that
	\begin{equation}
		\mathbb{E}[w(Y- s(\mathbf{t};\mathbf{\beta}) )^2] = \int_\mathcal{T}  \mathbb{E}[Y(\textbf{t})- s(\mathbf{t};\mathbf{\beta}) ]^2f_\textbf{T}(\textbf{t})d\textbf{t}.
	\end{equation}
	The proof of Proposition 1 can be found in Appendix A.2. Note that $Y(\textbf{t})$ on the right hand side of Equation (13) represents the potential outcome and $Y$ on the left hand side represents the observed outcome. Proposition 1 indicates that by having $w$ on the left hand side of Equation (13), one can represent the objective function with the potential outcome (right side) by that with the observed outcome (left side). Therefore, the true value $\beta^*$ is also a solution of the weighted optimization problem:
	\begin{equation}
		\beta^* = \text{argmin}_\beta  \mathbb{E}[w(Y- s(\mathbf{t};\mathbf{\beta}))^2].
	\end{equation}
	This result implies that the true value $\beta^*$ can be identified from the observational data.
	One can obtain the estimator based on the sample, which is
	\begin{equation}
		\hat{\beta} = \text{argmin}_\beta \sum_{i=1}^{n} \hat{w_i}(Y_i- s(\mathbf{T}_i;\mathbf{\beta}) )^2.
	\end{equation}
	
	\subsubsection{Nonparametric approach}
	Suppose $\mathbb{E}(Y(\mathbf{t})) = s(\mathbf{t})$. In a similar manner to the proof of Proposition 1, it can be shown that 
	\begin{equation*}
		\mathbb{E}(wY \mid \mathbf{T}=\textbf{t}) = \mathbb{E}(Y(\mathbf{t})).
	\end{equation*}
	Existing work of nonparametric tensor regression suffers from a slow rate of convergence due to the curse of dimensionality. Even if one flattens the tensor covariate into a vector and applies common nonparametric regression models such as additive models or single-index models to it, this issue still exists. Besides, when dealing with a vectorized tensor covariate, one would ignore the latent tensor structure and this might result in large bias. To meet these challenges, we adopt the broadcasted nonparametric tensor regression method (\citet{zhou2020broadcasted}) to estimate the causal effect function $s(\mathbf{t})$.
	\noindent
	
	The main idea of the broadcasted nonparametric tensor regression is to use the (low-rank) tensor structure to discover important regions of the tensor so as to broadcast a nonparametric modeling on such regions. Specifically, assume that 
	\begin{equation}
		s(\mathbf{T}) = c+\frac{1}{pq} \sum_{r=1}^{R}<\mathbf{\beta}_1^{(r)} \circ \mathbf{\beta}_2^{(r)}, F_r(\mathbf{T})>,
	\end{equation}
	where $c\in R, \ \mathbf{T} \in R^{p \times q}, \ F_r(\mathbf{T}) = \mathcal{B}(f_r, \mathbf{X})$, $\mathcal{B}$ is a broadcasting operator, which is defined as
	\begin{equation*}
		(\mathcal{B}(f,\mathbf{T}))_{i_1,i_2} = f(T_{i_1,i_2}),\ \text{ for all} \ i_1,i_2.
	\end{equation*}
	The broadcasted functions $f_r, r=1,\dots,R$, will be approximated by B-spline functions, i.e.,
	\begin{equation}
		f_r(x) \approx \sum_{d=1}^{D} \alpha_{r,d}b_d(x),
	\end{equation}
	where $\mathbf{b}(x) = (b_1(x),\dots,b_D(x))^{'}$ is a vector of B-spline basis functions and $\alpha_{r,d}$'s are the corresponding spline coefficients. Define $\mathbf{\alpha}_r = (\alpha_{r,1},\dots,\alpha_{r,D})^{'}$ and $(\Phi(\mathbf{T}))_{i_1,i_2,d} = b_d(T_{i_1,i_2})$, the regression function (16) can be approximated by 
	\begin{equation}
		s(\mathbf{T}) \approx c+\frac{1}{pq} \sum_{i=1}^{R}<\mathbf{\beta}_1^{(r)} \circ \mathbf{\beta}_2^{(r)} \circ \mathbf{\alpha}_r, \Phi(\mathbf{T})>.
	\end{equation}
	To separate out the constant effect from $f_r$'s, the condition $\int_{0}^{1} f_r(x) dx=0$ is imposed, which leads to 
	\begin{equation}
		\int_{0}^{1} \sum_{d=1}^{D} \alpha_{r,d}b_d(x)dx=0, \ r=1,\dots, R.
	\end{equation}
	Then the following optimization problem is considered:
	\begin{gather*}
		\text{argmin}_{c,\mathbf{G} }\ \sum_{i=1}^{n} (\hat{w}_iY_i- c-\frac{1}{pq}<\mathbf{G},\Phi(\mathbf{T}_i)>)^2
	\end{gather*}
	s.t.
	\begin{gather}
		\mathbf{G} = \sum_{r=1}^{R} \mathbf{\beta}_1^{(r)} \circ \mathbf{\beta}_2^{(r)} \circ \mathbf{\alpha}_r  \\
		\sum_{d=1}^{D} \alpha_{r,d}\int_{0}^{1} b_d(x)dx =0, \ r=1,\dots,R, \notag
	\end{gather}
	and the estimated regression function is 
	\begin{equation}
		\hat{s}(\mathbf{T}) = \hat{c}+\frac{1}{pq} <\hat{\mathbf{G}},\Phi(\mathbf{T})>,
	\end{equation}
	where $(\hat{c},\hat{\mathbf{G}})$ is a solution of (20). 
	\noindent
	
	Since optimization problem (20) contains too many constraints, it is not computationally efficient to solve it directly. To further simplify the optimization problem, an equivalent truncated power basis (\citet{ruppert2003semiparametric}) is used to reduce the constraints. Specifically, let $\tilde{b}_d(x), d=1,\dots,D$ denote the truncated basis:
	\begin{gather*}
		\tilde{b}_1(x)=1,\tilde{b}_2(x)=x,\dots, \tilde{b}_\varsigma(x) = x^{\varsigma-1},\\
		\tilde{b}_{\varsigma+1}(x) = (x-\xi_2)_{+}^{\varsigma-1},\dots, \tilde{b}_D(x)=(x-\xi_{D-\varsigma+1})_{+}^{\varsigma-1},
	\end{gather*}
	where $\varsigma$ and $(\xi_2,\dots,\xi_{D-\varsigma+1})$ are the order and the interior knots of the aforementioned B-spline, respectively. Based on these basis functions, consider the following optimization
	\begin{gather*}
		\text{argmin}_{\tilde{c},\tilde{\mathbf{G} }}\ \sum_{i=1}^{n} (\hat{w}_iY_i- \tilde{c}-\frac{1}{pq}<\tilde{\mathbf{G} },\tilde{\Phi}(\mathbf{T}_i)>)^2
	\end{gather*}
	s.t.
	\begin{gather}
		\tilde{\mathbf{G} } = \sum_{r=1}^{R} \mathbf{\beta}_1^{(r)} \circ \mathbf{\beta}_2^{(r)} \circ \tilde{\mathbf{\alpha}}_r,  	
	\end{gather}
	where $\tilde{\Phi}(\mathbf{T})_{i_1,i_2,d} = \tilde{b}_{d+1}(\mathbf{T}_{i_1,i_2}),\ d=1,\dots,D$ and $\tilde{\mathbf{\alpha}}_r \in R^{D-1}$ is the vector of coefficients. Compared with (20), the mean zero constraints are removed by reducing one degree of freedom in the basis functions. According to \cite{zhou2020broadcasted}, Lemma B.1, one can show that
	\begin{equation}
		\hat{s}(\mathbf{T}) = \hat{\tilde{c}}+\frac{1}{pq} <\hat{\tilde{\mathbf{G}}},\tilde{\Phi}(\mathbf{T})>,
	\end{equation}
	where $(\hat{\tilde{c}},\hat{\tilde{\mathbf{G}}})$ is the solution of (22). 
	\noindent
	The optimization problem (22) can be solved by the scaled-adjusted block-wise descent algorithm (\citet{zhou2020broadcasted}).
	
	\section{Theoretical properties}
	In this section, the large sample properties of the proposed estimators in section 3 are established. First the consistency of the estimated weight in section 3.1 is shown, then the consistency of the parametric estimator in section 3.2.1 and the convergence rate of the nonparametric estimator in section 3.2.2 are shown. The following assumptions are made. 
	\\
	\noindent
	\textbf{Assumption 4}
	\renewcommand \theenumi{\roman{enumi}}
	\renewcommand \labelenumi{(\theenumi)}
	\begin{enumerate}
		%	\item There exist two posistive constants $a_1$ and $a_2$ such that
		%	\begin{equation*}
		%		0< a_1 \leq w^* \leq a_2 < \infty.
		%	\end{equation*}
		\item There exists a constant $c_0$ such that $0 < c_0 < 1$, and $c_0 \leq \text{exp} (z-1) \leq 1-c_0$ for any $z= \tilde{M}_K(\mathbf{t},\mathbf{x})^{'}\mathbf{\theta}$ with $\mathbf{\theta} \in  \text{int}(\Theta)$. Besides, $\text{exp} (z-1) = O(1)$ in some neighborhood of $z^{*} = \tilde{M}_K(\mathbf{t},\mathbf{x})^{'}\theta^{*}$, where $\tilde{M}_K(\mathbf{t},\mathbf{x})= \Lambda (m_K(\mathbf{t},\mathbf{x})-\bar{m}_K)$ and $\Lambda= diag(\lambda_1,\dots,\lambda_K)$.
		\item There exists a constant $C$ such that %$\text{sup}_{(\mathbf{t},\mathbf{x})} \mid \mid M_K(\mathbf{t},\mathbf{x}) \mid \mid_2 \leq CK^{1/2}$ and 
		$E \lbrace \tilde{M}_K(\mathbf{T}_i, \mathbf{X}_i)\tilde{M}_K(\mathbf{T}_i, \mathbf{X}_i)^{'} \rbrace \leq C$. 
		\item $\delta = o(n)$.
		\item $	\text{sup}_{(\mathbf{T},\mathbf{X})} \  \text{exp}  \lbrace \sum_{j=1}^{K}\theta_j^* \lambda_j [m_{K,j}(\mathbf{T},\mathbf{X})-Em_{K,j}(\mathbf{T},\mathbf{X})] \rbrace = O(1)$.
	\end{enumerate}

	\noindent
	\textbf{Assumption 5}
	\begin{enumerate}
		\item The parameter space $\Theta_1$ is a compact set and the true parameter $\beta_0$ is in the interior of $\Theta_1$.
		\item $(Y-s(\textbf{T};\beta))^2$ is continuous in $\beta$, $\mathbb{E}[\text{sup}_\beta(Y-s(\textbf{T};\beta))^2] < \infty $ and $\text{sup}_\beta \mathbb{E}[(Y-s(\textbf{T};\beta))^4]  < \infty $.
	\end{enumerate}

	\noindent 
	\textbf{Assumption \ 6}
	\begin{enumerate}
		\item $s(\mathbf{t};\beta)$ is twice continuously differentiable in $\beta \in \Theta_1$ and let $h(\mathbf{t};\beta) \equiv \bigtriangledown_\beta s(\mathbf{t};\beta)$.
		\item $\mathbb{E} \lbrace w(Y-s(\mathbf{T};\beta))h(\mathbf{T};\beta) \rbrace $ is differentiable with respect to $\beta$ and \\
		$U \equiv - \bigtriangledown_\beta \mathbb{E} \lbrace w(Y-s(\mathbf{T};\beta))h(\mathbf{T};\beta) \rbrace \mid_{\beta=\beta^*}$ is nonsingular.
		\item $\mathbb{E}[\text{sup}_\beta \mid Y-s(\textbf{T};\beta) \mid^{2+\delta}] < \infty $ for some $\delta >0$ and there extists some finite positive constants $a$ and $b$ such that $\mathbb{E}[\text{sup}_{\beta_1:  \mid \mid \beta_1-\beta \mid \mid  < \delta_1} \mid s(\textbf{T};\beta_1) - s(\textbf{T};\beta) \mid^{2}]^{1/2} < a\cdot \delta_1^b $ for any $\beta \in \Theta_1$ and any small $\delta_1 >0$.
	\end{enumerate}

	\noindent
	\textbf{Assumption 7}
	\begin{enumerate}
		\item The treatment $\mathbf{T} \in [0,1]^{p\times q}$ has a continuous probability density function $f$, which is bounded away from zero and infinity.
		\item  The vector of random errors, $\mathbf{\epsilon} = (\epsilon_1,\dots,\epsilon_n)^{'}$, has independent and identically distributed entries. Each $\epsilon_i$ is sub-Gaussian with mean $0$ and sub-Gaussian norm $\sigma < \infty$.
		\item The true broadcasted functions $f_{0r} \in \mathcal{H}, r= 1,\dots,R_0$. Here $\mathcal{H}$ is the space of functions from [0,1] to $R$ satisfying the H$\ddot{o}$lder condition of order $\omega$, i.e.,
		\begin{equation*}
			\mathcal{H} = \lbrace g: \mid g^{(l)} (x_1)-g^{(l)} (x_2) \mid \leq S_1\mid x_1-x_2 \mid^{\omega}, \forall \ x_1,x_2 \in [0,1] \rbrace, 
		\end{equation*}
		for some constant $S_1>0$, where $g^{(l)}$ is the $l$-th derivative of $g$, such that $\omega \in (0,1]$ and $\tau = l+\omega >1/2$.
		\item The order of the $B$-spline used in (16) satisfies $\varsigma \geq \tau+1/2$. Let $0= \xi_1 < \xi_2 <\dots < \xi_{D-\varsigma+2}=1$ denote the knots of B-spline basis and assume that
		\begin{equation*}
			h_n = \text{max}_{d=1,\dots, D-\varsigma+1} \mid \xi_{d+1}
			-\xi_d \mid \asymp D^{-1} \ \text{and} \ \frac{h_n}{min_{d=1,\dots, D-\varsigma+1} \mid \xi_{d+1}
				-\xi_d \mid} \leq S_2 
		\end{equation*}
		for some constant $S_2>0$.
	\end{enumerate} 
	
	Assumption 4(\romannumeral1) enables consistency of $\hat{\theta}$ to translate into consistency of the weights.  Assumption 4(\romannumeral2) is a standard technical condition that restricts the magnitude of the basis functions; see also Assumption 4.1.6 of \citet{fan2016improving} and Assumption 2(\romannumeral2) of \citet{newey1997convergence}. Assumption 4(\romannumeral3) requires that the threshold parameter $\delta$ should be much smaller than the sample size. Assumption 4 (\romannumeral4) is needed for consistency of the estimated weight. Assumption 5(\romannumeral1) is a commonly used assumption in nonparametric regression. Assumption 5(\romannumeral2) is an envelope condition applicable to the uniform law of large numbers. Assumption 6(\romannumeral1) and (\romannumeral2)  impose sufficient regularity conditions on the causal effect function and its derivative function. Assumption 6(\romannumeral3) is a stochastic equicontinuity condition, which is needed for establishing weak convergence (\citet{1994Chapter}). Assumption 7(\romannumeral1), (\romannumeral3) and (\romannumeral4) are common in nonparametric regression models. In particular, Assumption 7(\romannumeral3) and (\romannumeral4) regularize the space where the true broadcasted functions lie in and guarantee that they can be approximated welll by B-spline functions. Similar assumptions can be found in \citet{shen1998local} and \citet{ huang2010variable}. Assumption 7(\romannumeral2) is a standard tail condition of the error. Based on these assumptions, the following theorems are established. \\
	\\
	
	\noindent 
	\textbf{Theorem 2.} \
	Let $\hat{\mathbf{\theta}}$ denote the solution to Problem (10) and 
	\begin{equation*}
		\hat{w}_i = \text{exp} \lbrace \sum_{k=1}^{K} \hat{\theta}_k \lambda_k (m_{K,k}(\mathbf{T}_i,\mathbf{X}_i)-\bar{m}_{K,k}) -1\rbrace , \ i=1,\dots,n,
	\end{equation*}
	then under Assumptions 1-4,\\
	\begin{enumerate}
		%	\item $\text{sup}_{(\mathbf{t},\mathbf{x})} \mid \hat{w} - w \mid = O_p(n^{-{1/2}}) $. 
		\item $\int \mid  \hat{\mathbf{w}}-\mathbf{w}^* \mid^2 dF(\mathbf{t},\mathbf{x}) = O_p(n^{-1})$.
		\item $\frac{1}{n} \sum_{i=1}^{n} \mid \hat{w}_i- w_i^* \mid^2 = O_p(n^{-1})$.
	\end{enumerate}
	\noindent
	
	Based on Theorem 2, we can establish the consistency of the parametric estimator and the convergence rate of the nonparametric estimator.\\
	
	\noindent 
	\textbf{Theorem 3} \ 
	\begin{enumerate}
		\item Under Assumptions 1-5, $\mid\mid \hat{\beta}-\beta^* \mid\mid \ \to_p 0$. 
		\item Under Assumptions 1-6, $\sqrt{n}(\hat{\beta}-\beta^*) \to_d N(0,V)$, where 
		\begin{equation*}
			V = 4U^{-1}\cdot \mathbb{E} \lbrace w^2(Y-s(\mathbf{T};\beta^*))^2h(\mathbf{T};\beta^*)h(\mathbf{T};\beta^*)^{'} \rbrace\cdot U^{-1} 
		\end{equation*}
	\end{enumerate}
	
	\noindent 
	\textbf{Theorem 4} \ If Assumptions 1-4 and 7 hold, $R\geq R_0$, and 
	\begin{equation*}
		n > S_1 h_n^{-2-2/\text{log}(h_n)}(\text{log}^{-2}(h_n))(R^{3}+R(p+q)+RD)
	\end{equation*}
	for some large enough constant $S_1>0$, then 
	%\begin{equation*}
	%	\begin{split}
	%	\mid\mid \hat{s}(\mathbf{T})-s(\mathbf{T}) \mid\mid^2 &= O_p(\lbrace \frac{(R^3+R(p+q)+RD)}{nD} \rbrace^{1/2}) 
	%	+O_p(\frac{\sum_{r=1}^{R_0}\mid\mid \text{vec}(\mathbf{B}_{0r})\mid\mid_1}{pq}\frac{1}{D^{\tau+1/2}}).
	%	\end{split}
	%\end{equation*}
	
	\begin{equation*}
		\mid\mid \hat{s}(\mathbf{T})-s_0(\mathbf{T}) \mid\mid^2 = O_p \bigg ( \frac{R^3+R(p+q)+RD}{n} \bigg )+O_p \bigg ( \big \lbrace \frac{\sum_{r=1}^{R_0} \mid\mid \text{vec}(\mathbf{B}_{0r})\mid\mid_1}{pq} \big \rbrace^2 \frac{1}{D^{2\tau}} \bigg ),
	\end{equation*}
	where $s_0(\mathbf{T}) = c_0+\frac{1}{pq} \sum_{r=1}^{R_0}<\mathbf{\beta}_1^{(0r)} \circ \mathbf{\beta}_2^{(0r)}, F_{0r}(\mathbf{T})>$ represents the true regression function.
	\noindent
	
	The proofs of Theorem 2, 3 and 4 can be found in Appendix A.3, A.4 and A.5, respectively.

	\section{Simulation}
	To evaluate the finite sample performance of the proposed method, simulation studies are carried out under different data settings. The main motivation of the simulation is to compare the proposed method with three other methods when the outcome model are linear and nonlinear in various ways.
	\subsection{The low-dimensional covariate setting}
	In this subsection, we compare the performance of the proposed method (WEBM) with the unweighted method (Unweighted), entropy balancing method (EB) and univariate approximate balancing method (MDABW) in the low-dimensional covariate setting, where EB refers to the method proposed by \citet{ai2021unified} that balances covariates exactly and MDABW refers to the method proposed by \citet{wang2020minimal} that balances covariates approximately.
	\noindent
	
	Since the covariates are shared across all scenarios, their data generating process is first described. Specifically, we independently draw 5 covariates from a multivariate normal distribution with mean 0, variance 1 and covariance 0.2, that is,
	
	\[ \mathbf{X} = (X_1,....,X_5)^{'} \sim  N_5(\mu,\Sigma) \ \text{with}\ \mu= \begin{pmatrix} 0\\  \vdots \\0 \end{pmatrix} \text{and}\ \Sigma=\begin{pmatrix} 1&0.2 &\dots& 0.2 \\0.2&1 &0.2 \dots & 0.2 \\  \hdotsfor{4} \\0.2&0.2&\dots &1 \end{pmatrix}_{5\times5.} \]
	\vspace{12pt}
	
	Consider a linear treatment assignment model, which is defined as
	\begin{equation*}
		\mathbf{T}_i= X_{i1}\mathbf{B}_1+X_{i2}\mathbf{B}_2+X_{i3}\mathbf{B}_3+\mathbf{E}_i,
	\end{equation*}
	where $\mathbf{B}_j = \begin{pmatrix} 1&0 \\ 0& 1 \\1&1 \end{pmatrix}_{3\times2}, j=1,2,3$ denotes the $j$th coefficient matrix, and $\mathbf{E}_i \in R^{3\times 2}$ denotes the error matrix, whose element follows a standard normal distribution. For the outcome model, we consider four scenarios and conduct 100 Monte Carlo simulations for each scenario. The first two scenarios assume an outcome model that is linear in treatment and the others assume a nonlinear relationship.
	\noindent
	
	In scenario 1, the linear outcome model is defined as
	\begin{equation}
		Y_i = 1+ <\mathbf{B}, \mathbf{T}_i>+X_{i1}+(X_{i2}+1)^2+X_{i4}^2+\epsilon_i,
	\end{equation}
	where $\mathbf{B} = \begin{pmatrix} 1&0 \\ 0& 1 \\1&1 \end{pmatrix}_{3\times2}$ and $\epsilon_i \sim N(0,2^2)$. In this scenario, $v_{k2}(\mathbf{X}) = (1, \mathbf{X}, \mathbf{X}*\mathbf{X})^{'}$, where $*$ represents the Hadamard product of two matrices, and the corresponding element of $\mathbf{X}*\mathbf{X}$ is $(\textbf{X}*\textbf{X})_{ij} = (x_{ij}x_{ij})$.
	
	\noindent
	
	In scenario 2, the linear outcome model is defined as
	\begin{equation}
		Y_i = 1+ <\mathbf{B}, \mathbf{T}_i>+X_{i2}+X_{i3}+X_{i1}X_{i2}+\dots+X_{i4}X_{i5}+\epsilon_i,
	\end{equation}
	where $\mathbf{B}$ and $\epsilon_i $ are the same as in Equation (21). In this scenario, the interaction terms are strong confounders, hence set $v_{k2}(\mathbf{X}) = (1, \mathbf{X}, X_jX_k)^{'}, 1\leq j < k \leq 5$.  
	
	In scenarios 3 and 4, the nonlinear outcome models are considered and are defined as
	\begin{equation}
		Y_i = 1+ <\mathbf{B}, F_1(\mathbf{T}_i)>+X_{i1}+(X_{i2}+1)^2+X_{i4}^2+\epsilon_i,
	\end{equation}
	and
	\begin{equation}
		Y_i = 1+ <\mathbf{B}, F_1(\mathbf{T}_i)>+X_{i2}+X_{i3}+X_{i1}X_{i2}+\dots+X_{i4}X_{i5}+\epsilon_i,
	\end{equation}
	respectively.
	Here $\mathbf{B}$ and $\epsilon_i$ are the same as in Equation (24), $(F_1(\mathbf{T}))_{k_1,k_2}= f_1(T_{k_1,k_2})= T_{k_1,k_2}+0.6sin\lbrace2\pi(T_{k_1,k_2}-0.5)^2 \rbrace$. For all four scenarios, consider $u_{K1}(\mathbf{T}) =(1, \text{vec}(\mathbf{T})^{'})^{'}$ for simplicity.
	\noindent
	
	For each method, the mean RMSE and its standard deviation of the coefficient estimates for the linear outcome model, and those of the fitted values for the nonlinear outcome model are reported based on 100 data replications. 
	
	Table 1 shows the mean RMSE and its standard deviation of the coefficient matrix for the linear outcome model. Observe that the mean RMSE of WEBM is the smallest among the four methods, and the standard deviation of WEBM is the second smallest while that of Unweighted is the smallest in both scenario 1 and scenario 2. Besides, the results of scenario 2 indicate that MDABW and EB  have poor performance when the basis function of covariates includes interaction entries, and their mean RMSEs are even larger than those of Unweighted. The mean RMSE and standard deviation of all methods decreases as the sample size increases.
	\noindent
	
	Table 2 shows the mean RMSE and its standard deviation of the fitted values of the nonlinear outcome model. As can be seen, the results for both scenario 3 and 4 are similar to those in Table 1, that is, WEBM performs the best with both the smallest mean RMSE in all cases. Similarly, MDABW and EB methods have poor performance when the basis function of covariates includes interaction entries and the mean RMSE and standard deviation of all methods decreases as the sample size increases.

	\begin{table}[t!]
		\begin{center}
			\caption{Tensor regression for the linear outcome model. Reported are the mean RMSE and its standard  deviation of the coefficient matrix based on 100 data replicates.}
			\label{tableone}
			\renewcommand \arraystretch{1.5}
			\resizebox{\linewidth}{!}{\begin{tabular}{cccccc}
					\hline
					&  \multicolumn{2}{c}{ \textbf{Scenario 1}} & &  \multicolumn{2}{c}{ \textbf{Scenario 2}} \\
					& $n=500$ & $n=1000$& & $n=500$ & $n=1000$ \\
					\textbf{Unweighted} & 0.6115 (\textbf{0.0192}) &0.6048 (\textbf{0.0129}) & & 0.5620 (\textbf{0.0238})  &0.5569 (\textbf{0.0145})   \\
					
					\textbf{MDABW} & 0.5357 (0.0983)&0.4601 (0.0447)& & 0.6477 (0.1425)& 0.6276 (0.1306) \\
					\textbf{EB} & 0.5442 (0.1068) &0.4615 (0.0469)  & & 0.7354 (0.1755)&  0.6994 (0.1518) \\ 
					\textbf{WEBM} & \textbf{0.5353} (0.0980) &\textbf{0.4585} (0.0385) & & \textbf{0.5385} (0.0935)& \textbf{0.5069} (0.0495)  \\ \hline
					
					%			   \textbf{Unweighted} & 1.5023 (0.0511) &1.4767 (0.0353) & & 1.3950 (0.0605)  &1.3641 (0.0511)   \\
					%			   
					%			   \textbf{WEBM} & 1.3045 (0.2552) &1.1819 (0.1719) & & 1.2752 (0.1650) & 1.2378 (0.1114) \\
					%			   \textbf{MDABW} & 1.3049 (0.2555)&1.1822 (0.1722) & & 1.6653 (0.3143) & 1.6468  (0.1676)\\
					%			   \textbf{EB} & 1.3050 (0.2556) &1.1823 (0.1723) & & 1.8061 (0.4848)&  1.7231 (0.3104) \\ \hline
			\end{tabular}}
		\end{center}
	\end{table}
	
	%	\begin{table}[t!]
	%	\begin{center}
	%		%			\setlength{\abovecaptionskip}{0cm}
	%		%			\setlength{\belowcaptionskip}{0.2cm}
	%		\caption{Tensor regression for the linear outcome model. Reported are mean RMSE and its standard  deviation of coefficient matrix based on 10 data replicates.}
	%		%\renewcommand \arraystretch{1.5}
	%		\label{tableone}
	%		\renewcommand \arraystretch{1.5}
	%		\resizebox{\linewidth}{!}{\begin{tabular}{cccccc}
	%				\hline
	%			 &  \multicolumn{2}{c}{ \textbf{Scenario 1}} & &  \multicolumn{2}{c}{ \textbf{Scenario 2}} \\
	%			 & $n=500$ & $n=1000$& & $n=500$ & $n=1000$ \\
	%			  \textbf{Unweighted} & 0.6133 (0.0209) &0.6028 (0.0144) & & 0.5695 (0.0263)  &0.5569 (0.0247)   \\
	%			  
	%			   \textbf{WEBM} & 0.5326 (0.1042) &0.4825 (0.0702) & & 0.5206 (0.0674)& 0.5053 (0.0455)  \\
	%			   \textbf{MDABW} & 0.5329 (0.1044)&0.4829 (0.0707) & & 0.6799 (0.1283)& 0.6523 (0.0684) \\
	%			   \textbf{EB} & 0.5331 (0.1046) &0.4832 (0.0709) & & 0.7373 (0.1979)&  0.7034 (0.1294) \\ \hline
	%			   
	%%			   \textbf{Unweighted} & 1.5023 (0.0511) &1.4767 (0.0353) & & 1.3950 (0.0605)  &1.3641 (0.0511)   \\
	%%			   
	%%			   \textbf{WEBM} & 1.3045 (0.2552) &1.1819 (0.1719) & & 1.2752 (0.1650) & 1.2378 (0.1114) \\
	%%			   \textbf{MDABW} & 1.3049 (0.2555)&1.1822 (0.1722) & & 1.6653 (0.3143) & 1.6468  (0.1676)\\
	%%			   \textbf{EB} & 1.3050 (0.2556) &1.1823 (0.1723) & & 1.8061 (0.4848)&  1.7231 (0.3104) \\ \hline
	%		\end{tabular}}
	%	\end{center}
	%\end{table}
	\begin{table}[t!]
		\begin{center}
			\caption{Tensor regression for the nonlinear outcome model. Reported are the mean RMSE and its standard deviation of fitted values based on 100 data replicates.}
			\label{tableone}
			\renewcommand \arraystretch{1.5}
			\resizebox{\linewidth}{!}{\begin{tabular}{cccccc}
					\hline
					&  \multicolumn{2}{c}{ \textbf{Scenario 3}} & &  \multicolumn{2}{c}{ \textbf{Scenario 4}} \\
					& $n=500$ & $n=1000$& & $n=500$ & $n=1000$ \\
					\textbf{Unweighted} & 6.4630 (\textbf{0.0016}) &6.4344 (\textbf{0.0013}) & & 5.3297 (\textbf{0.0073}) & 5.1756 (\textbf{0.0066})\\
					
					\textbf{MDABW} & 6.3987 (0.1864) &5.6535 (0.0030) & & 4.1064  (0.0185) & 3.6503 (0.0244)\\
					\textbf{EB} & 6.3988 (0.1865) &5.6536 (0.0031) & & 4.6743 (0.0796)&  4.1089 (0.0277) \\ 
					\textbf{WEBM} & \textbf{6.3775} (0.1843) &\textbf{5.6521} (0.0029) & & \textbf{3.5668} (0.0252) & \textbf{3.4452} (0.0169) \\ \hline
					
					%			   \textbf{Unweighted} & 1.5023 (0.0511) &1.4767 (0.0353) & & 1.3950 (0.0605)  &1.3641 (0.0511)   \\
					%			   
					%			   \textbf{WEBM} & 1.3045 (0.2552) &1.1819 (0.1719) & & 1.2752 (0.1650) & 1.2378 (0.1114) \\
					%			   \textbf{MDABW} & 1.3049 (0.2555)&1.1822 (0.1722) & & 1.6653 (0.3143) & 1.6468  (0.1676)\\
					%			   \textbf{EB} & 1.3050 (0.2556) &1.1823 (0.1723) & & 1.8061 (0.4848)&  1.7231 (0.3104) \\ \hline
			\end{tabular}}
		\end{center}
	\end{table}
	
	%	\begin{table}[t!]
	%	\begin{center}
	%		%			\setlength{\abovecaptionskip}{0cm}
	%		%			\setlength{\belowcaptionskip}{0.2cm}
	%		\caption{Tensor regression for the nonlinear outcome model. Reported are mean RMSE and its standard deviation of fitted values based on 10 data replicates.}
	%		%\renewcommand \arraystretch{1.5}
	%		\label{tableone}
	%		\renewcommand \arraystretch{1.5}
	%		\resizebox{\linewidth}{!}{\begin{tabular}{cccccc}
	%				\hline
	%				&  \multicolumn{2}{c}{ \textbf{Scenario 3}} & &  \multicolumn{2}{c}{ \textbf{Scenario 4}} \\
	%				& $n=500$ & $n=1000$& & $n=500$ & $n=1000$ \\
	%				
	%				\textbf{WEBM} & 6.2269 (0.5346) &5.6858 (0.1057) & & 3.6019 (0.0963) & 3.2531 (0.0331) \\
	%				\textbf{MDABW} & 6.2442 (0.5410) &5.6872 (0.1060) & & 4.3439 (0.1108) & 3.9721  (0.0616)\\
	%				\textbf{EB} & 6.2461 (0.5417) &5.6874 (0.1062) & & 4.6365 (0.1447)&  4.1222 (0.0834) \\ \hline
	%				
	%				%			   \textbf{Unweighted} & 1.5023 (0.0511) &1.4767 (0.0353) & & 1.3950 (0.0605)  &1.3641 (0.0511)   \\
	%				%			   
	%				%			   \textbf{WEBM} & 1.3045 (0.2552) &1.1819 (0.1719) & & 1.2752 (0.1650) & 1.2378 (0.1114) \\
	%				%			   \textbf{MDABW} & 1.3049 (0.2555)&1.1822 (0.1722) & & 1.6653 (0.3143) & 1.6468  (0.1676)\\
	%				%			   \textbf{EB} & 1.3050 (0.2556) &1.1823 (0.1723) & & 1.8061 (0.4848)&  1.7231 (0.3104) \\ \hline
	%		\end{tabular}}
	%	\end{center}
	%\end{table}

	\subsection{The high-dimensional covariate setting}
	In this subsection, since the MDABW and EB methods can only deal with the low-dimensional covariate case, we compare the performance of the proposed method (WEBM) with the Unweighted method and the method (Mapping) proposed by \citet{yu2022mapping}, which selects a small important subset of covariates by a joint screening procedure, %first screens out a small important subseteof covariates and estimates the causal effect by assuming a linear outcome model
	in the high-dimensional covariate setting. Since the Mapping method can only deal with linear outcome model, these two methods are compared in the linear outcome model setting.
	\noindent
	
	For both methods, set the sample size $n=500$ and the dimension of $\mathbf{T}$ to be $3 \times 2$. Consider two scenarios (scenario 5-6) with the dimension of covariates  $L=49$ and $L=99$, respectively. The motivation for such settings is to consider the number of constraints $K < n$ (scenario 5) and $K>n$ (scenario 6), respectively. For the basis functions, $u_{K1}(\mathbf{T})= (1,vec(\mathbf{T})^{'})^{'}$ and $v_{K2}(\mathbf{X}) = (1,\mathbf{X}^{'})^{'}$ are considered.
	Additionally, covariates are drawn from a multivariate normal distribution with mean 0, variance 1 and covariance 0.2. For each scenario, consider a linear treatment assignment model, which is defined as
	\begin{equation*}
		\mathbf{T}_i= X_{i1}\mathbf{B}_1+X_{i2}\mathbf{B}_2+\dots+X_{i5}\mathbf{B}_5+\mathbf{E}_i,
	\end{equation*}
	where $\mathbf{B}_j = \begin{pmatrix} 1&0 \\ 0& 1 \\1&1 \end{pmatrix}_{3\times2}, j=1,2,3,4,5$ denotes the $j$th coefficient matrix, and $\mathbf{E}_i \in R^{3\times 2}$ denotes the error matrix, whose element follows a standard normal distribution. 
	Moreover, the linear outcome model is defined as
	\begin{equation}
		Y_i = 1+ <\mathbf{B}, \mathbf{T}_i>+X_{i1}+X_{i2}+\dots+X_{i5}+\epsilon_i,
	\end{equation}
	% and 
	% \begin{equation}
	% 	Y_i = 1+ <\mathbf{B}, F_1(\mathbf{T}_i)>+X_{i1}+X_{i2}+\dots+X_{i5}+\epsilon_i,
	% \end{equation}
	%respectively. 
	where $\mathbf{B}=\begin{pmatrix} 1&0 \\ 0& 1 \\1&1 \end{pmatrix}_{3\times2}$, $\epsilon_i \sim N(0,2^2)$.% and $(F_1(\mathbf{T}))_{k_1,k_2}= f_1(T_{k_1,k_2})= T_{k_1,k_2}+0.6sin\lbrace2\pi(T_{k_1,k_2}-0.5)^2 \rbrace$.
	\noindent
	
	For each method, the mean RMSE and its standard deviation of the tensor regression estimation for the linear outcome model are reported based on 100 data replications.  %and the mean RMSE and its standard deviation of the fitted tensor regression for the nonlinear outcome model based on 200 data replications. 
	Table 3 shows the mean RMSE and its standard deviation of coefficent matrix for the linear outcome model in the high-dimensional covariate setting. The results indicate that WEBM performs the best with the smallest mean RMSE in all cases.
	\begin{table}[t!]
		\begin{center}
			\caption{Tensor regression for the linear outcome model. Reported are the mean RMSE and its standard deviation of the coefficient matrix based on 100 data replicates.}
			\label{tableone}
			\scriptsize
			\renewcommand \arraystretch{1.5}
			\resizebox{\linewidth}{!}{\begin{tabular}{cccc}
					\hline
					&  \textbf{Scenario 5 (L=49)} & &  \textbf{Scenario 6 (L=99)}\\
					\textbf{Unweighted} & 0.6140 (0.0053) & & 0.6147 (0.0059)  \\
					
					%	\textbf{Mapping} & 0.4890 (0.0619) & &0.4760 (0.0619)  \\ 
					\textbf{Mapping} & 0.5153 (0.0607) & &0.5326 (0.0821)  \\  
					\textbf{WEBM} & 0.4760 (0.0683) & & 0.4951 (0.0896) \\ \hline
					
			\end{tabular}}
		\end{center}
	\end{table}
	
	%	\begin{table}[t!]
	%	\begin{center}
	%		%			\setlength{\abovecaptionskip}{0cm}
	%		%			\setlength{\belowcaptionskip}{0.2cm}
	%		\caption{Tensor regression for the linear outcome model. Reported are mean RMSE and its standard  deviation of coefficient matrix based on 10 data replicates.}
	%		%\renewcommand \arraystretch{1.5}
	%		\label{tableone}
	%		\renewcommand \arraystretch{1.5}
	%		\resizebox{\linewidth}{!}{\begin{tabular}{cccc}
	%				\hline
	%				&  \textbf{Scenario 5 (L=49)} & &  \textbf{Scenario 6 (L=99)}\\
	%				\textbf{Unweighted} & 0.6128 (0.0046) & & 0.6160 (0.0036)  \\
	%				
	%				\textbf{WEBM} & 0.4514 (0.0276) & & 0.4674 (0.0648) \\
	%				\textbf{Mapping} & 0.5260 (0.0472) & &0.5367 (0.0711)  \\ \hline
	%				
	%		\end{tabular}}
	%	\end{center}
	%\end{table}
	
	\section{Application}
	Intelligence Quotient (IQ) is based on biological attributes, most of which are inherited from parents. It mainly refers to a person's cognitive abilities, such as logical reasoning, pattern recognition, and short-term memory. Of course, due to genetic mutations, parents with average IQ may have offspring with superior intelligence, and vice versa. But IQ also has social attributes. Studies have found that IQ has obvious plasticity, and environmental factors affect IQ levels. From 1947 to 2002, the IQ level of developed countries rose steadily at a rate of 3$\%$ every 10 years, which is called the "Flynn effect." This effect has been repeatedly observed in various countries, various age groups, and a large number of different environments, and has become a reliable evidence that "environment affects IQ" (\citet{brooks2012social}). In particular, literatures have shown that taking training lessons, such as music, sports, chess and so on, can significantly enhance children's IQ (\citet{kaviani2014can}; \citet{bradley2016dual}; \citet{schellenberg2004music}; \citet{joseph2016chess}). 
	\noindent
	
	Unfortunately, there are some limitations of existing studies. First, existing studies have only analyzed the relevant effects of attending single training course at a fixed age on children's IQ. In practice, however, a child may attend different training courses at the same age and participation may also vary by age. Second, existing studies mainly investigated the impact of whether or not to attend training classes, ignoring the effect of class duration. Third, existing studies only analyzed the correlation between children's participation in training courses on children's IQ, while their causal relationship is much more of a concern. To overcome the aforementioned limitations, the proposed method is applied to a matrix treatment, which contains the structural information of children's participation in training courses, to investigate their causal impact on children's IQ.
	\noindent
	
	The data are obtained from the Brain Science Innovation Institute of East China Normal University's Child Brain Intelligence Enhancement Project, whose goal is to explore brain development and help improve brain power. The treatment we are interested in is a $2 \times 5 $ matrix about children's participation in training courses, whose rows represent age groups, including 3-6 years old and 6-9 years old, columns represent the types of training courses, including knowledge education (Chinese, mathematics and English, etc.), art (music, art, calligraphy, etc.), sports (swimming, ball games, etc.), hands-on practice (STEM, Lego, etc.) and thinking training (logical thinking, EQ education, attention, etc.), and each element of the matrix represents the number of hours of class per week. The outcome is children's IQ and the pre-treatment covariates include children's gender as well as parental education, which have been shown to be associated with both the treatment and outcome variables (\citet{dubow2009long}; \citet{neiss2000parental}; \citet{dickson2016early}; \citet{cianci2013influence}). A complete-case analysis is conducted with a sample of 103 participants.  
	\noindent
	
	Before estimating the causal effect, we first examine the covariate balancing of WEBM, MDABW and EB methods based on the WEIM statistics. The statistic WEIM defined in equation (8) is 0.1050 for the WEBM method, 0.3761 for the MDABW method and 0.9881 for the EB method, which implies that WEBM balances covariates well while EB does not. Assume a linear tensor outcome model and the bootstrap method with 200 replicates is used to obtain confidence intervals for the parameter estimates. The results are shown in Table 4.
	\noindent
	
	\begin{landscape}
		
		\begin{table}[t!]
			\begin{center}
				\resizebox{\linewidth}{!}{	\begin{threeparttable}
						\caption{Tensor regression for the linear training course model.}
						\label{tableone}
						\renewcommand \arraystretch{1.5}
						\begin{tabular}{ccccccc}
							\hline
							\multirow{2}*{\textbf{Method}}	&  \multirow{2}*{\textbf{Age Group}} & \multicolumn{5}{c}{\textbf{Trainging type}} \\
							~&~&knowledge education& art &sports & hands-on practice & thinking trainging \\
							\multirow{4}*{Unweighted} & \multirow{2}*{3-6}&-0.0283 &-0.0368 &0.1139 & -0.7435& 0.0824 \\ 
							~&~&(-0.0654,02537)&(-0.1824,0.1340)&(-0.1572,0.4827)&(-1.2081,0.0089)&(-0.3897,0.4482)\\
							~&\multirow{2}*{6-9}&0.0329 & 0.0713& -0.1425&\textbf{1.0196}* &0.0093 \\ 
							~&~&(-0.1100,0.1744)&(-0.1647,0.2545)&(-0.3766,0.1046)&(0.1390,1.9436)&(-0.3757,0.6672) \\

							\multirow{4}*{MDABW} & 	\multirow{2}*{3-6}&-0.0278 &-0.0479 &0.1256 &-0.6182 &0.0544  \\ 
							~&~&(-0.0621,0.2900)&(-0.2027,0.1920)&(-0.0398,0.7358)&(-1.3027,0.0621)&(-0.0187,0.7870) \\
							~&	\multirow{2}*{6-9}&0.0252 &0.0982 &-0.1386 &\textbf{0.8487}* &0.0024 \\ 
							~&~&(-0.1474,0.2125)&(-0.1552,0.3809)&(-0.4805,0.1312)&(0.0015,1.8751)&(-0.5237,0.5284)\\
							
							\multirow{4}*{EB} & 	\multirow{2}*{3-6}&0.1594 &-0.0035 &0.6109 &-0.6971 &0.5870  \\ 
							~&~&(-0.1829,0.3907)&(-0.4321,0.2866)&(-0.2070,1.2768)&(-1.9252,0.5214)&(-0.3167,0.9092)\\
							~&	\multirow{2}*{6-9}&-0.0347 &0.0160 &-0.4245 &0.7889 &0.1152 \\  
							~&~&(-0.2497,0.2410)&(-0.4107,0.5346)&(-0.7109,0.1561)&(-0.5602,2.0901)&(-0.7951,1.2611)  \\ 
							\multirow{4}*{WEBM} & 	\multirow{2}*{3-6}&-0.0071 &-0.0467 & 0.2416& -0.7327&0.4260  \\ 
							~&~&(-0.0659,0.2362)&(-0.1937,0.1270)&(-0.1310,0.4788)&(-1.1737,0.0441)&(-0.3861,0.5885) \\
							~&	\multirow{2}*{6-9}&0.0081 &0.1241 &-0.2138 &\textbf{0.7532}* &0.0611 \\ 
							~&~&(-0.1166,0.1663)&(-0.1318,0.2882)&(-0.3873,0.1086)&(0.0009,1.8318)&(-0.3625,0.6866)\\ \hline

						\end{tabular}
						\begin{tablenotes}
							\item "*" represents that the confidence interval of the estimation does not contain zero, which implies that the effect of the estimation is siginificant.
						\end{tablenotes}
						
				\end{threeparttable}}
			\end{center}
		\end{table}
		
	\end{landscape}
	\noindent
	
	Table 4 shows the estimated causal effects of the duration of attending different classes at different ages on children's IQ. It can be seen that most methods (except EB) suggest that the duration of attending hands-on practice courses at 6-9 years old has a siginificantly positive impact on children's IQ. This finding is not only consistent with previous findings that participation in hands-on practice classes can improve children's IQ (\citet{legoff2006long}; \citet{owens2008lego}; \citet{sharifi2016lego}),  but further suggests that longer participation in hands-on practice classes is more beneficial to children's IQ.
	%which is consistent with the conslusions of the previous studies (\citet{legoff2006long}; \citet{owens2008lego}; \citet{sharifi2016lego}). 
	The results imply that future work of an intervention study about attending hands-on practice training courses, which in turn may improve children's IQ, is suggested. Besides, the width of confidence interval of the estimated causal effect based on WEBM is the smallest, which implies that the estimation accuracy of WEBM is the highest. 
	
	%Although the other estimations are not significant, all methods yielded consistent conclusions. It can be seen that all methods show opposite effects on children's IQ by attending different classes (except thinking training classes) at different ages. Specifically, most methods (except EB method) show a negative impact of the duration of taking knolwedge education courses on children's IQ in 3-6 years old while a postive impact of that in 6-9 yeas old. Similarly, all methods also show a negative impact of the duration of taking art and hands-on practice courses on children's IQ in 3-6 years old while a postive impact of that in 6-9 yeas old. Conversly, all methods show that the the longer the time spent in sports classes is benificial to children's IQ in 3-6 years old and is detrimental to children's IQ in 6-9 years old. Moreover, All methods suggest that the duration of taking thinking training classes have a positive impact on children's IQ at all different age groups. Furthermore, all methods agree that the duration of attending hands-on practice courses on children's IQ has the greatest impact on children's IQ at all ages group since their effect estimation is the largest.

	\section{Conclusion and discussion}	
	In this study, the weighted Euclidean balancing method is proposed, which obtains stabilized weights by adopting a single measure that represents the overall imbalance. An algorithm for the high-dimensional covariate setting is also proposed. Furthermore, parametric and nonparametric methods are developed to estimate the causal effect and their theoretical properties are provided. The simulation results show that the proposed method balances covariates well and produces a smaller mean RMSE compared to other methods under variaous scenarios. In the real data analysis, the WEBM method is applied to investigate causal effect of children's participation in training courses on their IQ. The results show that the duration of attending hands-on practice at 6-9 years old has a siginificantly positive impact on children's IQ.
	\noindent
	
	Since the causal effect function $\mathbb{E}(Y(t))$ is more general, we mainly consider it as the estimand for matrix treatment in this paper. Actually, one can also consider the average treatment effect  ($\mathbb{E}(Y(t+\triangle t)-Y(t))$) or average partial effect ($\frac{\mathbb{E}Y(t+\triangle t)-\mathbb{E}Y(t)}{\triangle t}$)  , which can be easily estimated based on the estimates of causal effect function  (\citet{2008A}). Indeed, the causal effect function provides a complete description of the causal effect, rather than a summary measure. Moreover, parametric and nonparametric methods are developed to estimate the causal effect function. Parametric method is recommended when reasonable assumptions can be made about the true model since it is easier to implement and requires less sample size. Despite nonparametric method has higher requirements of the sample size, one can choose to use it according to the real situations due to its higher flexibility. Besides, this paper mainly focuses on the small-scale matrix treatment. Large-scale matrix treatment with low-rank structure can also be considered. In such case, one may control the overall imbalance by only balancing their non-zero elements based on some decomposition technology, and this will be investigated in future work.

	\newpage
	\nocite{*}
	\bibliographystyle{apalike}
	\bibliography{ref}
	
	\newpage
	\section*{Appendix}
	\subsection*{A.1.\enspace Proof of Theorem 1} 
	The primal problem is
	\begin{gather*}
		\text{min}_\mathbf{w} \sum_{i=1}^{n}w_ilog(w_i)  \notag                                   
	\end{gather*}
	s.t.
	\begin{gather*}
		\sum_{j=1}^{K} \lbrace \lambda_j^2 [\sum_{i=1}^{n} w_i (m_{K,j}(\mathbf{T}_i,\mathbf{X}_i)-\bar{m}_{K,j})]^2 \rbrace \leq \delta.
	\end{gather*}
	
	Let $\mid\mid \theta \mid\mid_2 = \sqrt{\theta_1^2+\dots+\theta_K^2}$ be the $l_2$ norm for an arbitrary $K$-dimensional vector $\theta =(\theta_1,\dots,\theta_K)^{'}$ and $\Lambda = diag(\lambda_1,\dots,\lambda_K)$, then the inequality constraint in the primal problem can be rewritten as $\mid \mid \sum_{i=1}^{n} w_i \Lambda (m_K(\mathbf{T}_i,\mathbf{X}_i)-\bar{m}_K) \mid\mid_2 \leq \sqrt{\delta}$. Let $\mathcal{A} \subseteq R^K$ be a convex set such that $\mathcal{A} = \lbrace a \in R^K: \mid\mid a\mid\mid_2 \leq \sqrt{\delta} \rbrace$. Define $I_\mathcal{A}(a) = 0$ if $a \in \mathcal{A}$ and $I_\mathcal{A}(a) = \infty$ otherwise. Then, the primal problem (15) is equivalent to the following optimaization problem:
	\begin{equation*}
		\text{min}_\mathbf{w} \ \sum_{i=1}^{n}w_ilog(w_i)+I_\mathcal{A}( \sum_{i=1}^{n} w_i \Lambda (m_K(\mathbf{T}_i,\mathbf{X}_i)-\bar{m}_K)).
	\end{equation*} 
	\noindent
	
	Let $h(w) = \sum_{i=1}^{n} w_i log(w_i)$, the conjugate function of $h$ is 
	\begin{equation*}
		\begin{split}
			h^{*}(w) &= sup_t(\sum_{i=1}^{n} w_i t_i-\sum_{i=1}^{n} w_i log(w_i)) \\
			&= sup_t \sum_{i=1}^{n} (w_i t_i-w_i log(w_i)) \\
			&= \sum_{i=1}^{n} sup_{t_i}(w_i t_i-w_i log(w_i)) \\
			&=  \sum_{i=1}^{n} f^{*}(w_i),
		\end{split}
	\end{equation*}
	where $f^{*}(w_i) = \text{sup}_{t_i}(w_i t_i-w_i log(w_i))$ is the conjugate function of $f(w_i) = w_i log(w_i)$. Let $g(\theta) = I_\mathcal{A}(\theta)$ for any $\theta \in R^K$, then the conjugate function of $g$ is 
	\begin{equation*}
		\begin{split}
			g^{*}(\theta) &= \text{sup}_a (\sum_{k=1}^{K} \theta_ka_k -T_\mathcal{A}(a) ) \\
			&= \text{sup}_{\mid\mid a \mid\mid_2 \leq \sqrt{\delta}} (\sum_{k=1}^{K} \theta_ka_k) \\
			&= \text{sup}_{\mid\mid a \mid\mid_2 \leq \sqrt{\delta}}  (\mid\mid \theta \mid\mid_2 \mid\mid a \mid\mid_2) \\
			&= \sqrt{\delta} \mid\mid \theta \mid\mid_2.
		\end{split}
	\end{equation*}
	Define the mapping $H: R^n \to R^K$ such that $Hw = \sum_{i=1}^{n} w_i \Lambda (m_K(\mathbf{T}_i,\mathbf{X}_i)-\bar{m}_K)$, then $H$ is a bounded linear map. Let $H^{*}$ be the adjoint operator of $H$, then for all $\theta = (\theta_1,\dots,\theta_K)^{'} \in R^K$,
	\begin{equation*}
		H^{*}\theta = (\sum_{k=1}^{K}\theta_k \lambda_k (m_{K,k}(\mathbf{T}_1,\mathbf{X}_1)-\bar{m}_{K,k}), \dots, \sum_{k=1}^{K}\theta_k \lambda_k (m_{K,k}(\mathbf{T}_n,\mathbf{X}_n)-\bar{m}_{K,k}) )^{'}.
	\end{equation*}
	\noindent
	
	Define $\tilde{\theta} = H\tilde{w} = \frac{1}{n^r}\sum_{i=1}^{n} \Lambda (m_K(\mathbf{T}_i,\mathbf{X}_i)-\bar{m}_K)$, where $\tilde{w} = (\frac{1}{n^r},\dots,\frac{1}{n^r})^{'} \in dom(F)$. Here, we choose $b$ to be sufficiently large such that $\mid\mid \tilde{\theta} \mid\mid_2 \leq \sqrt{\delta}$, then we obtain that $g(\tilde{\theta}) = 0$ and $g$ is continuous at $\tilde{\theta}$. Therefore, $\tilde{\theta} \in H(dom(F) \cap cont(g))$, which implies that $H(dom(F) \cap cont(g)) \neq \emptyset$. Here, $dom(F)$ and $cont(g)$ denotes the domain of $F$ and the continuous set of $g$, respectively. Therefore, the strong duality condition of the Fenchel duality theorem is verified. Moreover,
	\begin{equation*}
		F(H^{*}\theta)+g^{*}(-\theta) = \sum_{i=1}^{n} f^{*}(\sum_{k=1}^{K}\theta_k \lambda_k (m_{K,k}(\mathbf{T}_i,\mathbf{X}_i)-\bar{m}_{K,k}))+\sqrt{\delta} \mid\mid \theta \mid\mid_2.
	\end{equation*}
	According to the Fenchel duality theorem (Mohri et al. (2018), Theorem B.39), we have 
	\begin{equation*}
		\begin{split}
			&\text{min}_w \ \sum_{i=1}^{n}w_ilog(w_i)+I_\mathcal{A}( \sum_{i=1}^{n} w_i \Lambda (m_K(\mathbf{T}_i,\mathbf{X}_i)-\bar{m}_K)) \\
			&= \text{min}_w \ \sum_{i=1}^{n} f^{*}(\sum_{k=1}^{K}\theta_k \lambda_k (m_{K,k}(\mathbf{T}_i,\mathbf{X}_i)-\bar{m}_{K,k}))+\sqrt{\delta} \mid\mid \theta \mid\mid_2.
		\end{split}
	\end{equation*} 
	Furthermore, since the strong duality condition holds, we can conclude that $H^{*}\hat{\theta}$ is a subgradient of $F$ at $\hat{w}$. That is, 
	\begin{equation*}
		\sum_{k=1}^{K}\hat{\theta}_k \lambda_k (m_{K,k}(\mathbf{T}_i,\mathbf{X}_i)-\bar{m}_{K,k}) = log(\hat{w}_i)+1.
	\end{equation*}
	Therefore, $\hat{w}_i = \text{exp}(\sum_{k=1}^{K}\hat{\theta}_k \lambda_k (m_{K,k}(\mathbf{T}_i,\mathbf{X}_i)-\bar{m}_{K,k})-1)$. The proof of theorem 1 is completed.

	\subsection*{A.2.\enspace Proof of Proposition 1}
	Using the law of total expectation and Assumption 1, we can deduce that
	\begin{gather*}
		\begin{split}
			&	\mathbb{E}[w(Y- s(\mathbf{T};\beta) )^2]\\
			&=E[\frac{f(\textbf{T})}{f(\textbf{T}\mid\textbf{X})}(Y-s(\mathbf{T};\beta)  )^2]\\
			&= \mathbb{E}(\mathbb{E}[\frac{f(\textbf{T})}{f(\textbf{T}\mid\textbf{X})}(Y- s(\mathbf{T};\beta)  )^2] \mid \textbf{T}=\textbf{t},\textbf{X}=\textbf{x} )\\
			&= \mathbb{E}(\frac{f(\textbf{t})}{f(\textbf{t}\mid\textbf{x})}\mathbb{E}([(Y- s(\mathbf{T};\beta)  )^2] \mid \textbf{T}=\textbf{t},\textbf{X}=\textbf{x}) )\\
			&= \int_{\mathcal{T}\times \mathcal{X}}\frac{f(\textbf{t})}{f(\textbf{t}\mid \textbf{x})}\mathbb{E}[(Y(\textbf{T})- s(\mathbf{T};\beta)  )^2 \mid \textbf{T} = \textbf{t}, \textbf{X}= \textbf{x}]f(\textbf{t}\mid \textbf{x})d\textbf{t}d\textbf{x}\\
			&=\int_{\mathcal{T}\times \mathcal{X}}\mathbb{E}[(Y(\textbf{T})- s(\mathbf{T};\beta)  )^2 \mid \textbf{T} = \textbf{t}, \textbf{X}= \textbf{x}]f(\textbf{t})f(\textbf{x})d\textbf{t}d\textbf{x}\\
			&= \int_{\mathcal{T}\times \mathcal{X}}\mathbb{E}[(Y(\textbf{t})- s(\mathbf{T};\beta) )^2 \mid \textbf{X}= \textbf{x}]f(\textbf{t})f(\textbf{x})d\textbf{t}d\textbf{x} \quad (\text{using Assumption 1})\\
			&= \int_{\mathcal{T}}\mathbb{E}[(Y(\textbf{t})- s(\mathbf{T};\beta)  )^2] f(\textbf{t})d\textbf{t}  .
		\end{split}
	\end{gather*}
	Hence, we complete the proof of Proposition 1.
	
	\subsection*{A.3.\enspace Proof of Theorem 2}
	The first order optimality condition for the dual probblem (10) is 
	\begin{equation*}
		\sum_{i=1}^{n} \text{exp} \lbrace\sum_{j=1}^{K} \hat{\theta}_j\lambda_jM_{K,j}(\mathbf{T}_i,\mathbf{X}_i) \rbrace \cdot \lambda_jM_{K,j}(\mathbf{T}_i,\mathbf{X}_i) +\sqrt{\delta} \frac{\hat{\theta}_j}{\mid\mid \hat{\theta} }\mid\mid_2 =0, \  \ j=1,\dots,K,
	\end{equation*}
	where $M_{K,j}(\mathbf{T}_i,\mathbf{X}_i) = m_{K,j}(\mathbf{T}_i,\mathbf{X}_i)- \bar{m}_{K,j}, M_{K}(\mathbf{T}_i,\mathbf{X}_i)= (M_{K,1}(\mathbf{T}_i,\mathbf{X}_i), \dots, M_{K,K}(\mathbf{T}_i,\mathbf{X}_i))^{'}$.
	Let $\Lambda = diag(\lambda_1,\dots,\lambda_K)$ and 
	\begin{equation*}
		\frac{1}{n}\sum_{i=1}^{n}\Phi(\mathbf{T}_i,\mathbf{X}_i;\theta) = \frac{1}{n} \sum_{i=1}^{n}\text{exp} \lbrace\sum_{j=1}^{K} \theta_j\lambda_j[m_{K,j}(\mathbf{T}_i,\mathbf{X}_i)-\mathbb{E}(m_{K,j})] \rbrace \Lambda [m_{K}(\mathbf{T}_i,\mathbf{X}_i) -\mathbb{E}(m_{K})],
	\end{equation*}
	which is a set of $K$ estimating functions. Note that
	\begin{equation*}
		\begin{split}
			&	\mid \mathbb{E}( \Phi(\mathbf{T}_i,\mathbf{X}_i;\theta^*)) \mid \\
			&= \mid \mathbb{E}
			\lbrace \text{exp} \lbrace\sum_{j=1}^{K} \theta_j^*\lambda_j[m_{K,j}(\mathbf{T}_i,\mathbf{X}_i)-Em_{K,j}] \rbrace \Lambda [m_{K}(\mathbf{T}_i,\mathbf{X}_i) -Em_{K}] \rbrace \mid \\
			&\leq \text{sup}_{(\mathbf{T}_i,\mathbf{X}_i)} \  \text{exp} \lbrace\sum_{j=1}^{K} \theta_j^*\lambda_j[m_{K,j}(\mathbf{T}_i,\mathbf{X}_i)-Em_{K,j}] \rbrace \cdot \mid \mathbb{E} \Lambda [m_{K}(\mathbf{T}_i,\mathbf{X}_i) -Em_{K}] \mid \\
			& \leq O(1) \cdot \mathbf{0} = \mathbf{0},
		\end{split}
	\end{equation*}
	hence we have $\mathbb{E}( \Phi(\mathbf{T}_i,\mathbf{X}_i;\theta^*))=\mathbf{0}$, which implies that $\theta^*$ is the unique solution of $\mathbb{E}( \Phi(\mathbf{T}_i,\mathbf{X}_i;\theta))=\mathbf{0}$. Therefore, by the estimating equation theory (Van der Vaart (2000)), the solution of the estimating equations 
	\begin{equation*}
		\frac{1}{n}\sum_{i=1}^{n}\Phi(\mathbf{T}_i,\mathbf{X}_i;\theta) = \mathbf{0},
	\end{equation*} 
	denoted by $\tilde{\theta}$, is asymptotically consistent for $\theta^*$. Furthermore, by the Taylor expansion, we have
	\begin{equation*}
		\sqrt{n}(\tilde{\theta} - \theta^*) \to_d N(\mathbf{0}, \Sigma),
	\end{equation*}
	where $\Sigma = \lbrace E(\frac{\partial \Phi}{\partial \theta^{'}}) \rbrace^{-1} E(\Phi\Phi^{'})\lbrace E(\frac{\partial \Phi}{\partial \theta}) \rbrace^{-1}$.
	Moreover, by the assumption that $\delta = o(n)$, we have $\frac{1}{n} \sqrt{\delta} \frac{\theta_j}{\mid\mid \theta \mid\mid_2} = o_p(n^{-1/2}) $ for any $\theta \in int(\Theta)$. Therefore, by the Slutsky's theorem, we obtain that 
	\begin{equation*}
		\sqrt{n}(\hat{\theta} - \theta^*) \to_d N(\mathbf{0}, \Sigma).
	\end{equation*}
	Let $\tilde{M}_{K,j}(\mathbf{T}_i,\mathbf{X}_i) = \lambda_j(m_{K,j}(\mathbf{T}_i,\mathbf{X}_i)- \bar{m}_{K,j})$,
	then 
	\begin{equation*}
		\begin{split}
			\hat{w}_i &= \text{exp} \lbrace \sum_{j=1}^{K} \hat{\theta}_j \lambda_j (m_{K,j}(\mathbf{T}_i,\mathbf{X}_i)-\bar{m}_{K,j}) -1\rbrace \\
			&= \text{exp} \lbrace \tilde{M}_K(\mathbf{T}_i,\mathbf{X}_i)^{'} \hat{\theta} -1\rbrace
		\end{split}
	\end{equation*}
	% Since 
	% \begin{equation*}
	% \begin{split}
	% 		&\text{sup}_{(\mathbf{t},\mathbf{x})} \mid \hat{w} - w^* \mid \\
	% 		&= \text{sup}_{(\mathbf{t},\mathbf{x})} \mid \text{exp} \lbrace \tilde{M}_K(\mathbf{t},\mathbf{x})^{'} \hat{\theta} -1\rbrace - \text{exp} \lbrace \tilde{M}_K(\mathbf{t},\mathbf{x})^{'} \theta^* -1\rbrace \mid \\
	% 		&\leq \text{exp} \lbrace \tilde{M}_K(\mathbf{t},\mathbf{x})^{'} \theta_1 -1\rbrace \times  \text{sup}_{(\mathbf{t},\mathbf{x})}  \mid \tilde{M}_K(\mathbf{t},\mathbf{x})^{'}(\hat{\theta}-\theta^*) \mid	\\
	% 		&\leq O_p(1) \cdot \mid\mid \hat{\theta}-\theta^* \mid\mid \cdot \text{sup}_{(\mathbf{t},\mathbf{x})}  \mid \tilde{M}_K(\mathbf{t},\mathbf{x}) \mid \\
	% 		&\leq O_P(1) \cdot O_P(n^{-1/2}) \cdot O(K^{1/2}) = O_p(\sqrt{\frac{K}{n}})=o_p(1),
	% \end{split}
	% \end{equation*}
	%where $\theta_1$ lies between $\hat{\theta}$ and $\theta^*$.

	By Mean Value Theorem, we can deduce that
	\begin{equation*}
		\begin{split}
			&		\int \mid \hat{w}-w^{*} \mid^2dF(\mathbf{t},\mathbf{x}) \\
			&	\leq \text{sup}_{(\mathbf{t},\mathbf{x})} \ \mid  \text{exp} \lbrace \tilde{M}_K(\mathbf{t},\mathbf{x})^{'} \theta_1 -1\rbrace \mid^2 \times \int \mid  \tilde{M}_K(\mathbf{t},\mathbf{x})^{'}(\hat{\theta}-\theta^{*}) \mid^2 dF(\mathbf{t},\mathbf{x}) \\
			&	\leq O_p(1) \cdot \int \mid  \tilde{M}_K(\mathbf{t},\mathbf{x})^{'}(\hat{\theta}-\theta^{*}) \mid^2 dF(\mathbf{t},\mathbf{x}),
		\end{split}
	\end{equation*}
	where $\theta_1$ lies between $\hat{\theta}$ and $\theta^*$.
	Since 
	\begin{equation*}
		\begin{split}
			&	\int \mid  \tilde{M}_K(\mathbf{t},\mathbf{x})^{'}(\hat{\theta}-\theta^{*}) \mid^2 dF(\mathbf{t},\mathbf{x}) \\
			& 	= \int \tilde{M}_K(\mathbf{t},\mathbf{x})^{'}(\hat{\theta}-\theta^{*})(\hat{\theta}-\theta^{*})^{'}\tilde{M}_K(\mathbf{t},\mathbf{x}) dF(\mathbf{t},\mathbf{x})\\
			&= tr\lbrace (\hat{\theta}-\theta^{*})(\hat{\theta}-\theta^{*})^{'} \int \tilde{M}_K(\mathbf{t},\mathbf{x})\tilde{M}_K(\mathbf{t},\mathbf{x})^{'}dF(\mathbf{t},\mathbf{x}) \rbrace \\
			&\leq C tr\lbrace (\hat{\theta}-\theta^{*})(\hat{\theta}-\theta^{*})^{'} \rbrace \\
			&= C\mid \mid \hat{\theta}-\theta^{*} \mid\mid^2 \\
			&= O_p(n^{-1}).
		\end{split}
	\end{equation*}
	Then we have 	
	\begin{equation*}
		\int \mid \hat{w}-w^{*} \mid^2dF(\mathbf{t},\mathbf{x}) = O_p(n^{-1}).
	\end{equation*}
	Furthermore, one can show that 
	\begin{equation*}
		\begin{split}
			\frac{1}{n}\sum_{i=1}^{n} \mid \tilde{M}_K(\mathbf{t},\mathbf{x})^{'}(\hat{\delta}-\delta^{*}) \mid^2 - \int \mid  \tilde{M}_K(\mathbf{t},\mathbf{x})^{'}(\hat{\delta}-\delta^{*}) \mid^2 dF(\mathbf{t},\mathbf{x})=o_p(1).
		\end{split}
	\end{equation*}
	Hence,
	\begin{equation*}
		\begin{split}
			&\frac{1}{n} \sum_{i=1}^{n} \mid \hat{w}_i- w_i^{*} \mid^2 \\
			&\leq \text{sup}_{(\mathbf{t},\mathbf{x})} \ \mid \text{exp} \lbrace \tilde{M}_K(\mathbf{t},\mathbf{x})^{'} \theta_1 -1\rbrace  \mid^2 \cdot \frac{1}{n} \sum_{i=1}^{n} \mid \tilde{M}_K(\mathbf{t},\mathbf{x})^{'}(\hat{\theta}-\theta^{*}) \mid^2 \\
			& \leq O_p(1) \int \mid  \tilde{M}_K(\mathbf{t},\mathbf{x})^{'}(\hat{\theta}-\theta^{*}) \mid^2 dF(\mathbf{t},\mathbf{x}) +o_p(1) \\
			&= O_p(n^{-1}) 
		\end{split}
	\end{equation*}
	Therefore, the proof of Theorem 2 is completed.
	
	\subsubsection*{A.4.\enspace Proof of Theorem 3}

	We first show that the conclusion of Theorem 3(\romannumeral1). \\
	
	Since $\hat{\beta}$ (as a estimator of $\beta^*$) is a unique minimizer of $\frac{1}{n}\sum_{i=1}^{n}\hat{w}_i(Y_i-s(\textbf{T}_i;\beta))^2$(regarding $\mathbb{E}[w(Y-s(\textbf{T};\beta))^2]$, according to the theory of M-estimation (van der Vaart, 2000, Theorem 5.7), if%(van der Vaart, 2000, Theorem 5.7), 
	\begin{gather*}
		\text{sup}_{\beta\in \Theta_1} \mid \frac{1}{n}\sum_{i=1}^{n}\hat{w_i}(Y_i-s(\textbf{T}_i;\beta))^2-\mathbb{E}[w(Y-s(\textbf{T};\beta))^2]) \mid \to_p 0,
	\end{gather*}
	then $\hat{\beta} \to_p \beta^*$.
	Note that
	\begin{gather}
		\text{sup}_{\beta\in \Theta_1} \mid \frac{1}{n}\sum_{i=1}^{n}\hat{w_i}(Y_i-s(\textbf{T}_i;\beta))^2-\mathbb{E}[w(Y-s(\textbf{T};\beta))^2]) \mid  \notag \\
		\leq \text{sup}_{\beta\in \Theta_1} \mid \frac{1}{n}\sum_{i=1}^{n}(\hat{w_i}-w_i)(Y_i-s(\textbf{T}_i;\beta))^2 \mid \notag \\
		+\text{sup}_{\beta\in \Theta_1} \mid \frac{1}{n}\sum_{i=1}^{n}w_i(Y_i-s(\textbf{T}_i;\beta))^2-\mathbb{E}[w(Y-s(\textbf{T};\beta))^2]) \mid.
	\end{gather}
	We first show that $\text{sup}_{\beta\in \Theta_1} \mid \frac{1}{n}\sum_{i=1}^{n}(\hat{w_i}-w_i)(Y_i-s(\textbf{T}_i;\beta))^2 \mid$ is $o_p(1)$. Using the Causchy-Schwarz inequality and the fact that $\hat{w}\to^{L^2} w$, we have
	\begin{gather*}
		\begin{split}
			\text{sup}_{\beta\in \Theta_1} \mid \frac{1}{n}\sum_{i=1}^{n}(\hat{w_i}-w_i)(Y_i-s(\textbf{T}_i;\beta))^2 \mid  & \leq \lbrace  \frac{1}{n}\sum_{i=1}^{n}(\hat{w_i}-w_i)^2 \rbrace^{1/2} \text{sup}_{\beta\in \Theta_1} \lbrace \frac{1}{n}\sum_{i=1}^{n}(Y_i-s(\textbf{T}_i;\beta))^2 \rbrace^{1/2} \\
			& \leq o_p(1)\lbrace \text{sup}_{\beta\in \Theta_1} \mathbb{E}[w(Y-s(\textbf{T};\beta))^2]+o_p(1) \rbrace^{1/2}\\
			&=o_p(1).
		\end{split}
	\end{gather*}
	Thereafter, under Assumption 5, we can conclude that $\text{sup}_{\beta\in \Theta_1} \mid \frac{1}{n}\sum_{i=1}^{n}w_i(Y_i-s(\textbf{T}_i;\beta))^2-\mathbb{E}[w(Y-s(\textbf{T};\beta))^2]) \mid $ is also $o_p(1)$ (Newey and McFadden (1994), Lemma 2.4). Hence, we complete the proof for Theorem 3(\romannumeral1). Next, we give the proof of Theorem 3(\romannumeral2). Define
	\begin{equation*}
		\hat{\beta}^* = \text{argmin}_\beta \sum_{i=1}^{n} w_i(Y_i-s(\mathbf{T}_i;\beta))^2.
	\end{equation*}
	Assume that $\frac{1}{n} \sum_{i=1}^{n} w_i(Y_i - s(\mathbf{T}_i;\hat{\beta}^*))h(\mathbf{T}_i;\hat{\beta}^*)) = o_p(n^{-1/2})$ holds with probablility to one as $n \to \infty$
	
	By Assumption 5 and the uniform law of large number, one can get that
	\begin{equation*}
		\frac{1}{n}\sum_{i=1}^{n} w_i(Y_i-s(\textbf{T}_i;\beta))^2 \to \mathbb{E} \lbrace w(Y-s(\textbf{T};\beta))^2 \rbrace \  \text{in probability uniformly over } \ \beta,
	\end{equation*}
	which implies $\mid \mid \hat{\beta}^* -\beta^* \mid \mid \to_p 0$. Let
	\begin{equation*}
		r(\beta) = 2\mathbb{E} \lbrace w(Y-s(\textbf{T};\beta))h(\textbf{T};\beta) \rbrace,
	\end{equation*}
	which is a differentiable function in $\beta$ and $r(\beta^*) = 0$. By mean value theorem, we have
	\begin{equation*}
		\sqrt{n}r(\hat{\beta}^*)- \bigtriangledown_\beta r(\zeta) \cdot \sqrt(n)(\hat{\beta}^* - \beta^*) =\sqrt{n}r(\beta^*) =0
	\end{equation*}
	where $\zeta$ lies on the line joining $\hat{\beta}^*$ and $\beta^*$. Since $\bigtriangledown_\beta r(\beta)$ is continuous at $\beta^*$ and $\mid \mid \hat{\beta}^* -\beta^* \mid \mid \to_p 0$, then
	\begin{equation*}
		\sqrt{n}(\hat{\beta}^* - \beta^*)  = \bigtriangledown_\beta r(\beta^*)^{-1}\cdot \sqrt{n} r(\hat{\beta}^*) +o_p(1)
	\end{equation*}
	Define the empirical process
	\begin{equation*}
		G_n(\beta)= \frac{2}{\sqrt{n}} \sum_{i=1}^{n} \lbrace w_i(Y_i-s(\textbf{T}_i;\beta))h(\textbf{T}_i;\beta) - \mathbb{E} \lbrace w(Y-s(\textbf{T};\beta))h(\textbf{T};\beta)  \rbrace \rbrace.
	\end{equation*}
	Then we have
	\begin{equation*}
		\begin{split}
			&	\sqrt{n}(\hat{\beta}^* - \beta^*)\\
			&=  \bigtriangledown_\beta r(\beta^*)^{-1}\cdot \lbrace  \sqrt{n} r(\hat{\beta}^*) -  \frac{2}{\sqrt{n}} \sum_{i=1}^{n} \lbrace w_i(Y_i-s(\textbf{T}_i;\hat{\beta}^*))h(\textbf{T}_i;\hat{\beta}^*)  + \frac{2}{\sqrt{n}} \sum_{i=1}^{n} \lbrace w_i(Y_i-s(\textbf{T}_i;\hat{\beta}^*))h(\textbf{T}_i;\hat{\beta}^*)  \rbrace \\
			&=  -\bigtriangledown_\beta r(\beta^*)^{-1}\cdot G_n(\hat{\beta}^*)+o_p(1)\\
			&= U^{-1} \cdot \lbrace G_n(\hat{\beta}^*)-G_n(\beta^*) +G_n(\beta^*) \rbrace +o_p(1).
		\end{split}	
	\end{equation*}
	By Assumption 5, 6, Theorem 4 and 5 of Andrews(1994), we have $G_n(\hat{\beta}^*)-G_n(\beta^*) \to_p 0$. Thus,
	\begin{equation*}
		\sqrt{n}(\hat{\beta}^* - \beta^*) = U^{-1} \frac{2}{\sqrt{n}} \sum_{i=1}^{n} \lbrace w_i(Y_i-s(\textbf{T}_i;\beta^*))h(\textbf{T}_i;\beta^*) \rbrace +o_p(1),
	\end{equation*}
	then we can get that the asymptotic variance of $\sqrt{n}(\hat{\beta}^* - \beta^*)$ is $V$.
	Therefore, $\sqrt{n}(\hat{\beta}^* - \beta^*) \to_d N(0,V)$. Next, we will prove $\hat{\beta} \to_p \hat{\beta}^*$.
	Since
	
	\begin{gather*}
		\text{sup}_{\beta\in \Theta_1} \mid \frac{1}{n}\sum_{i=1}^{n}\hat{w_i}(Y_i-s(\textbf{T}_i;\beta))^2-\frac{1}{n}\sum_{i=1}^{n}w_i(Y_i-s(\textbf{T}_i;\beta))^2) \mid  \\
		\leq \text{sup}_{\beta\in \Theta_1} \mid \frac{1}{n}\sum_{i=1}^{n}(\hat{w_i}-w_i)(Y_i-s(\textbf{T}_i;\beta))^2 \mid  \\
		\leq \lbrace  \frac{1}{n}\sum_{i=1}^{n}(\hat{w_i}-w_i)^2 \rbrace^{1/2} \text{sup}_{\beta\in \Theta_1} \lbrace \frac{1}{n}\sum_{i=1}^{n}(Y_i-s(\textbf{T}_i;\beta))^2 \rbrace^{1/2} \\
		\leq o_p(1)\lbrace \text{sup}_{\beta\in \Theta_1} \mathbb{E}[w(Y-s(\textbf{T};\beta))^2]+o_p(1) \rbrace^{1/2}\\
		=o_p(1),
	\end{gather*}
	which implies $\hat{\beta}^* \to_p \hat{\beta}$. Then by Slutskey's Theorem, we can draw the conclusion that $\sqrt{n}(\hat{\beta} - \beta^*) \to_d N(0,V)$. Therefore, we have completed the proof of Theorem 3. 
	
	\subsubsection*{A.5.\enspace Proof of Theorem 4}
	For convenience, we use a mapping $\Omega: R^{p\times q \times D} \times R \to R^{p\times q \times D}$ to represent the operator of absorbing the constant into the coefficients of B-spline basis for the first predictor. More precisely, $\Omega$ is defined by 
	\begin{equation*}
		\mathbf{G}^b= \Omega(\mathbf{G},c),
	\end{equation*}
	where $\mathbf{G}_{i_1,i_2,d}= \mathbf{G}_{i_1,i_2,d}$ for $(i_1,i_2) \neq (1,1)$ and $\mathbf{G}_{1,1,d}=\mathbf{G}_{1,1,d}+pqc, d=1,\dots,D$. It then follows from the property of B-spline functions that
	\begin{equation*}
		c+\frac{1}{pq}<\mathbf{G},\Phi(\mathbf{T})> = \frac{1}{pq} <\mathbf{G}^b,\Phi(\mathbf{T})>.
	\end{equation*}
	We also write $\mathbf{G}_0= \sum_{r=1}^{R_0} \mathbf{B}_{0r} \circ \mathbf{\alpha}_{0r}, r= 1,\dots, R_0$. Suppose $\hat{\mathbf{G}},\hat{c})$ is a solution to (19) and 
	\begin{equation*}
		\hat{\mathbf{G}} = \sum_{r=1}^{R} \hat{\mathbf{\beta}}_1^{(r)} \circ \hat{\mathbf{\beta}}_2^{(r)} \circ \hat{\mathbf{\alpha}}_r,
	\end{equation*}
	then by \citet{zhou2020broadcasted}, Lemma B.1, there exists $\check{c} \in R$ and 
	\begin{equation*}
		\check{\mathbf{G}} = \sum_{r=1}^{R} \hat{\mathbf{\beta}}_1^{(r)} \circ \hat{\mathbf{\beta}}_2^{(r)} \circ \check{\mathbf{\alpha}}_r,
	\end{equation*}
	such that
	\begin{equation}
		\check{c}+\frac{1}{pq}<\check{\mathbf{G}} ,\Phi(\mathbf{T})> = \hat{c}+\frac{1}{pq}<\hat{\mathbf{G}} ,\tilde{\Phi}(\mathbf{T})>,
	\end{equation}
	where $\check{\mathbf{\alpha}}_r= (\check{\alpha}_{r,1},\dots,\check{\alpha}_{r,D} )^{'}$ satisfying
	\begin{equation*}
		\sum_{d=1}^{D}\check{\alpha}_{r,d}u_d=0
	\end{equation*}
	with $u_d= \int_{0}^{1} b_d(x)dx$.
	Using (27), %By the deifinition of $(\hat{\mathbf{G}},\hat{c})$, 
	we have
	\begin{equation}
		\sum_{i=1}^{n} (\hat{w}_iy_i-\check{c}-\frac{1}{pq}<\check{\mathbf{G}},\Phi(\mathbf{T}_i)>)^2 \leq \sum_{i=1}^{n} (\hat{w}_iy_i-c_0-\frac{1}{pq}<\mathbf{G}_0,\Phi(\mathbf{T}_i)>)^2.
	\end{equation} 
	Let $\check{\mathbf{G}}^b= \Omega(\check{\mathbf{\alpha}},\check{c})$ and $\mathbf{G}_0^b = \Omega(\mathbf{G}_0,c_0)$, then 
	\begin{equation}
		\sum_{i=1}^{n} (\hat{w}_iy_i-\frac{1}{pq}<\check{\mathbf{G}}^b,\Phi(\mathbf{T}_i)>)^2 \leq \sum_{i=1}^{n} (\hat{w}_iy_i-\frac{1}{pq}<\mathbf{G}_0^b,\Phi(\mathbf{T}_i)>)^2.
	\end{equation}
	Therefore, we have
	\begin{equation}
		\begin{split}
			\sum_{i=1}^{n} ((\hat{w}_i-w_i+w_i)y_i-\frac{1}{pq}<\check{\mathbf{G}}^b,\Phi(\mathbf{T}_i)>)^2 \leq \sum_{i=1}^{n} ((\hat{w}_i-w_i+w_i)y_i-\frac{1}{pq}<\mathbf{G}_0^b,\Phi(\mathbf{T}_i)>)^2, 
		\end{split}
	\end{equation}
	which leads to 
	\begin{equation}
		\begin{split}
			\sum_{i=1}^{n} (w_iy_i-\frac{1}{pq}<\check{\mathbf{G}}^b,\Phi(\mathbf{T}_i)>)^2 &\leq \sum_{i=1}^{n} (w_iy_i-\frac{1}{pq}<\mathbf{G}_0^b,\Phi(\mathbf{T}_i)>)^2 \\
			&+2 \sum_{i=1}^{n} (\hat{w}_i-w_i)y_i(\frac{1}{pq}<\check{\mathbf{G}}^b-\mathbf{G}_0,\Phi(\mathbf{T}_i)).
		\end{split}
	\end{equation}
	Let $\mathbf{G}^{\#} =\check{\mathbf{G}}^b-\mathbf{G}_0^b, \ \mathbf{a}^{\#} =\text{vec}(\mathbf{G}^{\#} ), \ \mathbf{a}_0^b =\text{vec}(\mathbf{G}_0^b),\ \check{\mathbf{a}}^b= \text{vec}(\check{\mathbf{G}}^b)$ and $\mathbf{Z}
	= (\mathbf{z}_1,\dots,\mathbf{z}_n)^{'} \in R^{n\times pqD}$, where $\mathbf{z}_i = \text{vec}(\Phi(\mathbf{T}_i)),i=1,\dots,n$. 
	Let $y_{\hat{w} }= (\hat{w}_1y_1,\dots, \hat{w}_ny_n)^{'}$ and $y_w = (w_1y_1,\dots, w_ny_n)^{'}$, then using (31) and working out the squares, we obtain
	\begin{equation}
		\begin{split}
			\frac{1}{p^2q^2} \mid\mid \mathbf{Z}\mathbf{a}^{\#} \mid\mid^2& \leq  2<\frac{1}{pq}\mathbf{Z}\check{\mathbf{a}}^b,y_{w}>-2<	\frac{1}{pq}\mathbf{Z}\mathbf{a}_0^b,y_{w}>\\
			&-2\frac{1}{p^2q^2}<\mathbf{Z}\mathbf{a}^{\#},\mathbf{Z}\mathbf{a}_0^b>+2<\frac{1}{pq} \mathbf{Z}\mathbf{a}^{\#}, y_{\hat{w}}-y_w>  \\
			&= 2<\frac{1}{pq}\mathbf{Z}\mathbf{a}^{\#},y_{\hat{w}}-y_w>+2<\frac{1}{pq}\mathbf{Z}\mathbf{a}^{\#},\mathbf{\epsilon}>+2<	\frac{1}{pq}\mathbf{Z}\mathbf{a}^{\#},y_{w}-\mathbf{\epsilon}-\frac{1}{pq}\mathbf{Z}\mathbf{a}_0^b> 
			%	&\leq nO_p(n^{-1})+2<\frac{1}{pq}\mathbf{Z}\check{\mathbf{a}}^b,y_{\hat{w}}-y_w>+2<\frac{1}{pq}\mathbf{Z}\mathbf{a}^{\#},\mathbf{\epsilon}>\\
			%	&+2<	\frac{1}{pq}\mathbf{Z}\mathbf{a}^{\#},y_{w}-\mathbf{\epsilon}-\frac{1}{pq}\mathbf{Z}\mathbf{a}_0^b> \\
			%	&= O_p(1)+2<\frac{1}{pq}\mathbf{Z}\check{\mathbf{a}}^b,y_{\hat{w}}-y_w>+2<\frac{1}{pq}\mathbf{Z}\mathbf{a}^{\#},\mathbf{\epsilon}>\\
			%	&+2<	\frac{1}{pq}\mathbf{Z}\mathbf{a}^{\#},y_{w}-\mathbf{\epsilon}-\frac{1}{pq}\mathbf{Z}\mathbf{a}_0^b>.
		\end{split}
	\end{equation}
	\noindent
	
	First, we show the upper bound of $<\frac{1}{pq}\mathbf{Z}\mathbf{a}^{\#},y_{\hat{w}}-y_w>$. Using the Cauchy-Schwarz inequality, we have
	\begin{equation}
		\begin{split}
			&<\frac{1}{pq}\mathbf{Z}\mathbf{a}^{\#},y_{\hat{w}}-y_w> \\
			&\leq   \mid\mid \hat{w}-w \mid\mid_2 \cdot \mid\mid \frac{1}{pq} \mathbf{Z}\mathbf{a}^{\#} \mid\mid_2 \\
			&\leq \frac{C_1\sqrt{nh_n}}{pq} \cdot \mid\mid \hat{w}-w \mid\mid_2 \cdot \mid\mid \mathbf{a}^{\#}\mid\mid_2 
		\end{split}
		%		&\leq O_p(\lbrace \frac{D(R^3+pqR+RD)}{n} \rbrace^{1/2})+O_p(\frac{\sum_{r=1}^{R_0}\mid\mid \text{vec}(\mathbf{B}_{0r})\mid\mid_1}{pq}\frac{1}{D^{\tau-1/2}})	\end{split}
	\end{equation}
	
	By the conclusion of Theorem 2(\romannumeral2), we have 
	\begin{equation}
		\mid\mid \hat{w}-w \mid\mid_2 = O_p(1).
	\end{equation}
	Applying (37) to (36), we can obtain that
	\begin{equation}
		\begin{split}
			<\frac{1}{pq}\mathbf{Z}\mathbf{a}^{\#},y_{\hat{w}}-y_w> \leq \frac{C_2\sqrt{nh_n}}{pq} \mid\mid \mathbf{a}^{\#} \mid\mid_2.
		\end{split}
	\end{equation}
	Second, by the conclusion of \citet{zhou2020broadcasted}, (A.18) and (A.20), we can obtain the upper bound of
	$	<\frac{1}{pq}\mathbf{Z}\mathbf{a}^{\#},\mathbf{\epsilon}>$ and $<	\frac{1}{pq}\mathbf{Z}\mathbf{a}^{\#},y_{w}-\mathbf{\epsilon}-\frac{1}{pq}\mathbf{Z}\mathbf{a}_0^b>$, which are
	\begin{equation}
		\begin{split}
			<\frac{1}{pq}\mathbf{Z}\mathbf{a}^{\#},\mathbf{\epsilon}> \leq \frac{C_3}{pq} \mid\mid \mathbf{a}^{\#} \mid\mid_2 \lbrace nh_n(R^3+R(p+q)+RD)\rbrace^{1/2}.
		\end{split}
	\end{equation}
	and
	\begin{equation}
		<\frac{1}{pq}\mathbf{Z}\mathbf{a}^{\#},y_{w}-\mathbf{\epsilon}-\frac{1}{pq}\mathbf{Z}\mathbf{a}_0^b> \leq \frac{C_4}{pq} \mid\mid \mathbf{a}^{\#} \mid\mid_2 \lbrace\frac{\sum_{r=1}^{R_0} \mid\mid \text{vec}(\mathbf{B}_{0r})\mid\mid_1}{pq} \rbrace \frac{n\sqrt{h_n}}{D^\tau}
	\end{equation}
	Therefore, applying (38), (39) and (40) to (35), we have
	\begin{equation}
		\begin{split}
			\frac{C_5}{pq} \mid\mid \mathbf{a}^{\#} \mid\mid_2^2 \leq R_1 \mid\mid \mathbf{a}^{\#} \mid\mid_2,
		\end{split}
	\end{equation}
	where  $R_1 = C_6\sqrt{\frac{D}{n}}+C_7  \lbrace \frac{D(R^3+R(p+q)+RD)}{n}\rbrace^{1/2}+C_8\lbrace\frac{\sum_{r=1}^{R_0} \mid\mid \text{vec}(\mathbf{B}_{0r})\mid\mid_1}{pq} \rbrace \frac{1}{D^{\tau-1/2}}$.
	
	By solving the second order inequality (41), we have
	\begin{equation}
		\frac{C_5}{pq} \mid\mid \mathbf{a}^{\#} \mid\mid_2 \leq R_1
	\end{equation}
	Further, by Assumption 6 and \cite{zhou2020broadcasted}, (A.38) of Lemma A.2, we have
	\begin{equation}
		\begin{split}
			\mid\mid \hat{s}(\mathbf{T}) -s(\mathbf{T}) \mid\mid^2& \leq C_9 h_n \frac{1}{p^2q^2} \mid\mid \mathbf{a}^{\#} \mid\mid^2 \\
			&= \frac{C_{10}R_1^2}{D} \\
			&= O_P(\frac{1}{n})+O_p(\frac{R^3+R(p+q)+RD}{n})+O_p(\lbrace\frac{\sum_{r=1}^{R_0} \mid\mid \text{vec}(\mathbf{B}_{0r})\mid\mid_1}{pq} \rbrace^2 \frac{1}{D^{2\tau}}) \\
			&=O_p(\frac{R^3+R(p+q)+RD}{n})+O_p(\lbrace\frac{\sum_{r=1}^{R_0} \mid\mid \text{vec}(\mathbf{B}_{0r})\mid\mid_1}{pq} \rbrace^2 \frac{1}{D^{2\tau}}).
		\end{split}
	\end{equation}
	Hence, the proof of Theorem 4 is completed. 
	
	\section*{Appendix Reference}
	Mohri, M., Rostamizadeh, A., and Talwalkar, A. (2018). \emph{Foundations of ma- chine learning}. MIT press.
	\vskip 0.2cm
	\noindent
	Newey, W. K. and McFadden, D. (1994). Large sample estimation and hypoth- esis testing. \emph{Handbook of econometrics}, 4:2111–2245.
	\vskip 0.2cm
	\noindent
	Van der Vaart, A. W. (2000). \emph{Asymptotic statistics}, volume 3. Cambridge university press.
	\vskip 0.2cm
	\noindent
	Zhou, Y., Wong, R., and He, K. (2020). Broadcasted nonparametric tensor regression.
\end{document}